\documentclass[twocolumn]{aastex631}

\usepackage{multirow}

\newcommand\xmm{\textit{XMM-Newton}}
\newcommand\chandra{\textit{Chandra}}

\newcommand\athena{\textit{NewAthena}}

\newcommand\delcstat{$\Delta$C-stat}
\newcommand\pcm{cm$^{-2}$}

\newcommand\logxi{$\log (\xi$/erg~cm~s$^{-1})$}

\shorttitle{Ionized outflow in GSN 069}
\shortauthors{Kosec et al.}
\graphicspath{{./}{figures/}}

\begin{document}

\title{Detection of a Highly Ionized Outflow in the Quasi-periodically Erupting Source GSN 069}

\correspondingauthor{P. Kosec}
\email{peter.kosec@cfa.harvard.edu}

\author[0000-0003-4511-8427]{P. Kosec}
\affiliation{MIT Kavli Institute for Astrophysics and Space Research, Massachusetts Institute of Technology, Cambridge, MA 02139}
\affiliation{Center for Astrophysics | Harvard \& Smithsonian, Cambridge, MA, USA}

\author[0000-0003-0172-0854]{E. Kara}
\affiliation{MIT Kavli Institute for Astrophysics and Space Research, Massachusetts Institute of Technology, Cambridge, MA 02139}

\author[0000-0003-2663-1954]{L. Brenneman}
\affiliation{Center for Astrophysics | Harvard \& Smithsonian, Cambridge, MA, USA}

\author[0000-0002-0568-6000]{J. Chakraborty}
\affiliation{MIT Kavli Institute for Astrophysics and Space Research, Massachusetts Institute of Technology, Cambridge, MA 02139}

\author[0000-0002-1329-658X]{M. Giustini}
\affiliation{Centro de Astrobiología (CAB), CSIC-INTA, Camino Bajo del Castillo s/n, 28692 Villanueva de la Cañada, Madrid, Spain}

\author[0000-0003-0707-4531]{G. Miniutti}
\affiliation{Centro de Astrobiología (CAB), CSIC-INTA, Camino Bajo del Castillo s/n, 28692 Villanueva de la Cañada, Madrid, Spain}

\author[0000-0003-2532-7379]{C. Pinto}
\affiliation{INAF - IASF Palermo, Via U. La Malfa 153, I-90146 Palermo, Italy}

\author[0000-0002-5359-9497]{D. Rogantini}
\affiliation{Department of Astronomy and Astrophysics, The University of Chicago, Chicago, IL 60637}

\author[0000-0003-4054-7978]{R. Arcodia}
\affiliation{MIT Kavli Institute for Astrophysics and Space Research, Massachusetts Institute of Technology, Cambridge, MA 02139}

\author{M. Middleton}
\affiliation{School of Physics \& Astronomy, University of Southampton, Southampton, Southampton SO17 1BJ, UK}

\author[0000-0002-7295-5661]{A. Sacchi}
\affiliation{Center for Astrophysics | Harvard \& Smithsonian, Cambridge, MA, USA}




\begin{abstract}

Quasi-periodic eruptions (QPEs) are high-amplitude, soft X-ray bursts recurring every few hours, associated with supermassive black holes. Many interpretations for QPEs were proposed since their recent discovery in 2019, including extreme mass ratio inspirals and accretion disk instabilities. But, as of today, their nature still remains debated. We perform the first high-resolution X-ray spectral study of a QPE source using the RGS gratings onboard \xmm, leveraging nearly 2~Ms of exposure on GSN~069, the first discovered source of this class. We resolve several absorption and emission lines including a strong line pair near the N VII rest-frame energy, resembling the P-Cygni profile. We apply photoionization spectral models and identify the absorption lines as an outflow blueshifted by $1700-2900$~km/s, with a column density of about $10^{22}$ \pcm\ and an ionization parameter \logxi\ of $3.9-4.6$. The emission lines are instead redshifted by up to 2900~km/s, and likely originate from the same outflow that imprints the absorption features, and covers the full $4\pi$ sky from the point of view of GSN~069. The column density and ionization are comparable to the outflows detected in some tidal disruption events, but this outflow is significantly faster and has a strong emission component. The outflow is more highly ionized when the system is in the phase during which QPEs are present, and from the limits we derive on its location, we conclude that the outflow is connected to the recent complex, transient activity of GSN~069 which began around 2010.

\end{abstract}



\section{Introduction} 
\label{sec:intro}

Quasi-periodic eruptions (hereafter QPEs) are recently discovered high-amplitude transient phenomena, first detected in GSN 069 \citep{Miniutti+19}, which appear to occur in some low-mass supermassive black holes ($10^5-10^7 $ M$_{\odot}$). They involve a sudden spike in the X-ray flux of the source, mostly in the soft X-ray band ($<2$ keV) with a blackbody spectral shape ($kT\sim100$ eV), a duration of $1-30$ ks \citep{Miniutti+23a}, and a repeating timescale from hours to tens of days, depending on the source. As of February 2024, 5 sources were known to exhibit QPEs in addition to GSN 069 \citep{Sun+13, Giustini+20, Arcodia+21, Arcodia+24}, and further systems are good candidates or likely related to the QPE phenomenon \citep{Chakraborty+21, Quintin+23, Evans+23, Guolo+24}. These sources show that QPEs are primarily an X-ray band phenomenon, and their host systems exhibit little, if any nuclear (supermassive black hole related) activity in other energy bands. Additionally, QPEs have been observed to be a transient phenomenon, appearing and disappearing for years at a time \citep[GSN 069,][]{Miniutti+23b}, as well as showing a long-term flux fading trend \citep[eRO-QPE1,][]{Chakraborty+24, Pasham+24b}.

In the last few years, many models have been proposed to explain QPEs. The phenomenon may be related to tidal disruption events and/or extreme mass ratio inspirals \citep{Sukova+21, Xian+21, King+22, Metzger+22, Krolik+22, Franchini+23, Lu+23, Linial+23, Tagawa+23, Zhou+24a, Zhou+24b}. Alternative models propose that it is connected to accretion disk instabilities or disk tearing \citep{Raj+21, Pan+22, Pan+23, Sniegowska+23, Kaur+23}. However, all of these models struggle to explain all the complex behaviour of QPE systems, and so the nature of QPEs remains to be determined. Crucially, more observational work is required to reveal further physics and the emission mechanism of these intriguing systems.

All previous X-ray spectral QPE studies exclusively focused on data from the moderate spectral resolution X-ray instruments such as the CCDs onboard \xmm\ and \chandra, which offer the highest effective area, and result in the best count rates. However, the very soft (kT$\sim50-100$ eV) blackbody-like spectral energy distribution (SED) of QPE sources both during their quiescence as well as QPEs makes them also ideal targets for X-ray gratings such as the Reflection Grating Spectrometers (RGS) onboard \xmm, which operate in this soft X-ray band ($0.35-2$ keV). RGS gratings offer a lower collecting area compared with the CCD instruments, and hence worse statistics, but much better spectral resolution. Therefore, with grating spectra, we may resolve individual spectral lines in these systems, revealing further the physics of the QPE phenomenon. As an example, high-spectral resolution X-ray grating spectra of two tidal disruption events \citep{Miller+15, Kosec+23b}, which have X-ray spectral shapes similar to those of QPE sources, revealed complex absorption line spectra imprinted by ionized outflows.

By the end of 2023, individual QPE sources have been the targets of intensive observational campaigns - e.g., \xmm\ has already observed GSN 069 for almost 2 Ms. Here we present the first in-depth high-spectral resolution study of a QPE system, making use of this extensive archive of RGS spectra on GSN 069. In Section \ref{sec:data}, we describe our data reduction and stacking procedures. In Section \ref{sec:results}, we present our spectral modelling approach of the RGS dataset and the main results of this study. In the following Section \ref{sec:epicresults}, we analyze the simultaneous \xmm\ European Photon Imaging Camera (EPIC) datasets. Finally, in Section \ref{sec:discussion} we discuss our findings and their implications and list the conclusions in Section \ref{sec:conclusions}.

\section{Data reduction and stacking} \label{sec:data}

We analyze all \xmm\ \citep{Jansen+01} observations of GSN 069 performed by the end of July 2023. These data were downloaded from the \xmm\ Science Archive and reduced using \textsc{sas v20}, \textsc{caldb} as of February 2023. We primarily work with data from the RGS \citep{denHerder+01}, but also reduce data from the EPIC pn and MOS instruments \citep{Struder+01, Turner+01}.

RGS data were filtered using standard methods with the \textsc{rgsproc} routine, while centering the extraction region position on the location of GSN 069. Any background flares were filtered with a threshold of 0.25 counts/s. We primarily work with observational background spectra, but also tested the robustness of our results using blank field backgrounds. Data from the two RGS instruments (RGS 1 and 2) were not stacked, instead we always analyze them simultaneously using an extra cross-calibration constant parameter. The value of this constant was always within 10\% of unity, indicating good agreement between RGS 1 and RGS 2. We binned the RGS spectra by a factor of 3 to achieve mild oversampling of the instrumental spectral resolution. This was achieved with the `bin' command in the \textsc{spex} fitting package \citep{Kaastra+96}. Wavelength range in which the RGS spectra are used varies from dataset to dataset. We use as broad wavelength band as possible, and only cut off any range which has no source signal. This often happens on the lower end of the RGS wavelength coverage (given the spectral softness of GSN 069). The upper end of the wavelength range is 36 \AA, on the edge of the fully calibrated RGS band. Additionally, we also removed the 31 to 33.5 \AA\ region in all RGS 1 datasets as there is a large spike in background flux at these wavelengths, greatly exceeding the source flux in all datasets. In this region we only use RGS 2 data.

EPIC-pn and EPIC-MOS data were processed alongside the reduced RGS datasets using the \textsc{epproc} and \textsc{emproc} routines, and filtered such that only events of PATTERN$\leq$4 were accepted for pn data, and only events of PATTERN$\leq$12 were accepted for MOS data. The data were screened for background flares using a threshold of 0.4 ct/s in the $10-12$ keV pn lightcurve. The source region for both instruments was a circle with a radius of 35 arcsec centered on the position of GSN 069. The background region was a polygon on the same CCD as the source, maximizing the region area while being at least 125 arcsec away from the source. Background-subtracted EPIC-pn lightcurves were then produced using the \textsc{epiclccorr} routine. The data were grouped using the \textsc{specgroup} procedure to at least 25 counts per bin and at the same time to oversample the instrumental resolution by at most a factor of 3. The EPIC-pn data are used between 0.3 and 4.0 keV, and the EPIC-MOS data are used between 0.5 keV and 4.0 keV.

\subsection{Stacking}

Individual RGS spectra of GSN 069 have poor data quality due to the faintness of the source. They have at most $\sim2000$ net source counts (RGS 1+2), but typically show only hundreds of net counts. Therefore, we make use of data stacking. Initially, we stacked the RGS 1 and RGS 2 spectra from all observations together into just two spectral files (RGS 1 stack and RGS 2 stack). We note that we never analyze RGS 1 and 2 data stacked into a single spectral file (such stacking is only done for plotting purposes). Instead, we always stack individual RGS exposures into one RGS 1 and one RGS 2 spectrum, which are analyzed simultaneously using a cross-calibration constant. This stacking procedure yielded two spectra totalling 12544 and 13451 net counts for RGS 1 and 2, respectively. This is a very high data quality achieved using a combined net exposure of about 1.6 Ms. These stacked spectra are analyzed in section \ref{sec:Gaussianscan}. While the source spectrum has many counts, as GSN 069 is very faint, background is comparable or stronger than source flux throughout all of the RGS wavelength range. For this reason, we perform further checks of the RGS background in Appendix \ref{app:RGSbkg}.

\begin{figure}
\begin{center}
\includegraphics[width=0.85\columnwidth]{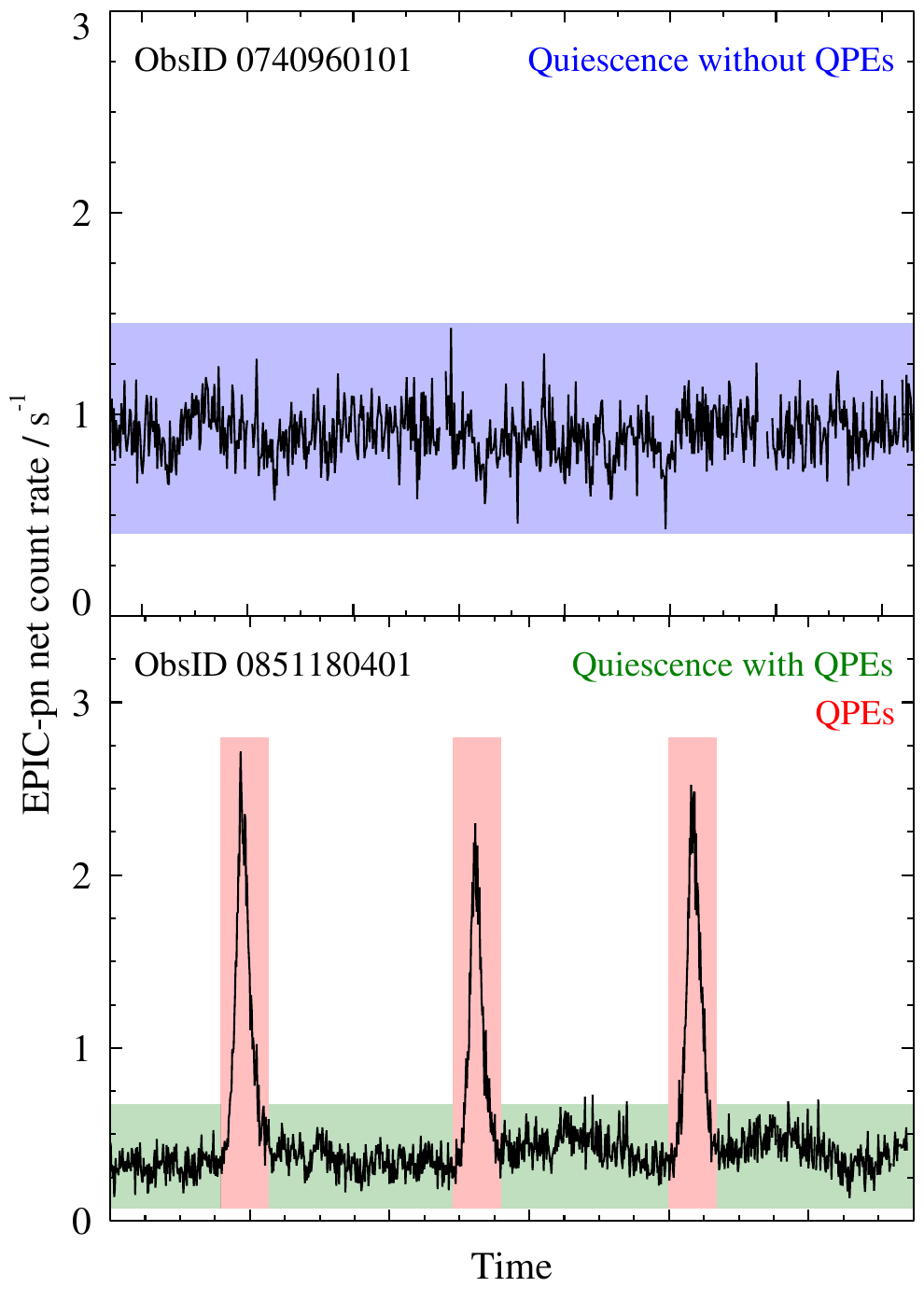}
\caption{EPIC-pn lightcurves from two example \xmm\ observations, showing how we denote the 3 states of GSN 069 in this work: Quiescence without QPEs, Quiescence with QPEs and the QPE state. \label{Example_lc}}
\end{center}
\end{figure}

Since there are over 25000 net RGS counts in total, we were able to do a data split. To track the time and state evolution of GSN 069, we split the dataset into 3 groups: 1. QPE state (time during QPEs), 2. Quiescent state with QPEs (time between individual QPEs during QPE active observations), 3. Quiescent state without QPEs (observations when the source is not showing any QPE activity). For clarity, we show these 3 states using EPIC-pn lightcurves of two example \xmm\ observations in Fig. \ref{Example_lc}.

To perform this split, we identified all observations in which QPEs were observed, primarily using EPIC-pn lightcurves, which were double-checked with RGS lightcurves. We identified the intervals of QPE activity during each observation (with a typical duration of $1-3$ ks) and saved them in good-time interval (GTI) files. Then, for these \xmm\ observations containing QPEs, we extracted the RGS QPE, and Quiescence spectra using these GTI files. Finally, we combined all data from the individual observations into these 3 state stacks. The same kind of splitting procedure was applied to the EPIC pn and MOS datasets, using the same GTI files.

We note that there are a number of observations performed between 30 November 2022 and 15 January 2023 with durations shorter than the typical time interval between two consecutive QPEs. 11 out of 19 of these observations contain QPEs. Here we assume that GSN 069 was constantly in the QPE active state during the period of time between Nov 2022 and Jan 2023, despite the remaining 8 \xmm\ observations not directly showing evidence for QPEs. We include these 8 observations in the Quiescent state with QPEs dataset stack.

The details of the stacked RGS datasets are given in Table \ref{RGS_data_info}. Further details of all individual \xmm\ observations are listed in Table \ref{xmmdata} in Appendix \ref{app:xmmdata}. The quiescence state spectra (with or without QPEs) both still have very good data quality (around 10000 net RGS counts each), while the QPE spectrum has poorer quality with about 5000 net RGS counts due to the lower net exposure time.

\begin{deluxetable*}{ccccc}
\tablecaption{Details of the RGS spectra in the 3 stacked datasets. \label{RGS_data_info}}
\tablewidth{0pt}
\tablehead{
\colhead{State name} & \multicolumn2c{RGS1} & \multicolumn2c{RGS2} \\
\nocolhead{} & \colhead{net exposure} & \colhead{net counts} & \colhead{net exposure} & \colhead{net counts}  \\
\nocolhead{} & \colhead{ks} & \colhead{} & \colhead{ks} & \colhead{}
}
\startdata 
Quiescence without QPEs  & 448.7 & 4955 & 452.9 & 5222 \\
Quiescence with QPEs  & 1042.5 & 5191 & 1054.3 & 5834 \\
QPE  & 108.5 & 2503 & 109.1 & 2401 \\
\enddata
\end{deluxetable*}

 \section{RGS dataset Spectral modelling and results} \label{sec:results}

All spectra are fitted in the \textsc{spex} (version 3.07.03) spectral fitting package \citep{Kaastra+96}. To analyze GSN 069 spectra using this package, we converted the reduced spectra from \textsc{ogip} format into \textsc{spex} format using the \textsc{trafo} routine. We quote all uncertainties at $1\sigma$ significance level, and use Cash statistic \citep{Cash+79} for spectral fitting. We adopt a redshift of $z=0.018$ for GSN 069 \citep{Jones+09}.

\subsection{Stacked RGS Grating Spectrum of GSN 069}
\label{sec:Gaussianscan}

As the first step of this analysis, we consider the stacked RGS spectrum of GSN 069 including all of the \xmm\ observations of this source. This is the highest signal-to-noise RGS spectrum, but necessarily it will also contain the most observation-to-observation variability as it contains all the states of GSN 069 - the QPEs, as well as the quiescence. To investigate the statistical significance of any tentative spectral lines in the stacked RGS spectrum, we perform a Gaussian line scan. 

First, the spectrum is fitted with a baseline continuum model. We chose a disk blackbody (\textsc{dbb} in \textsc{spex}) as the baseline emission model. We note here that the definition of the \textsc{dbb} temperature differs from the definition of temperature in the popular \textsc{diskbb} model in \textsc{xspec}, resulting in roughly a factor of 2 difference between the two values when fitted to the same spectrum\footnote{Further info on the \textsc{dbb} model can be found at https://spex-xray.github.io/spex-help/models/dbb.html}. We model the Galactic (nearly neutral) gas absorption using the \textsc{hot} model, where the gas temperature is set to $10^{-3}$ eV and its column density is fixed to $2.3 \times 10^{20}$ \pcm\ \citep{HI4PI+16}. Additionally, we use the \textsc{reds} model to take into account the redshift of GSN 069.

On top of this baseline continuum, we fit a Gaussian line which can have either positive or negative normalization (either emission or absorption line allowed), and has a fixed wavelength and width. The width is calculated from a pre-defined line velocity width, and the wavelength is varied according to a wavelength grid. We search the spectrum using two representative line widths - 100 km/s (a narrow line) and 1000 km/s (a moderately broadened line). We use a wavelength grid spanning the full energy range of the RGS spectrum (18.5 \AA\ to 36 \AA), spaced by 0.01 \AA. For each wavelength in the grid, we fit the emission model with the additional Gaussian line and recover the \delcstat\ fit improvement over the baseline continuum.

While this \delcstat\ value cannot be precisely converted into statistical significance of a potential line using a single Gaussian scan (without performing Monte Carlo simulations) as we did not account for the look-elsewhere effect, \delcstat\ is still a good indicator of whether a feature is real. For comparison, we highlight recent work on ultraluminous X-ray sources (ULXs) and AGN, where much computational effort was spent to quantify how different spectral fit improvements \delcstat\ (upon adding an outflow component to the baseline continuum model) translate to detection significance. Even though there is no universal rule for this relationship, typically, a \delcstat\ fit improvement of $19-22$ translates to a significance of about $3\sigma$ \citep{Pinto+20, Kosec+18a, Kosec+20b}. Conversely, a \delcstat\ fit improvement of about 30 usually translates to a significance of $4\sigma$ \citep{Kosec+18b, Pinto+21}. We further note that these ULX spectra, acquired also with \xmm\ RGS, typically have between 5000 and 30000 net counts (RGS 1 + 2 combined), which is very similar to our GSN 069 RGS datasets in this work. At least around 5000 net counts in the combined RGS spectrum are necessary to begin detecting line features at $3-4\sigma$ level \citep{Kosec+18a, Kosec+21}.

\begin{figure*}
\begin{center}
\includegraphics[width=\textwidth]{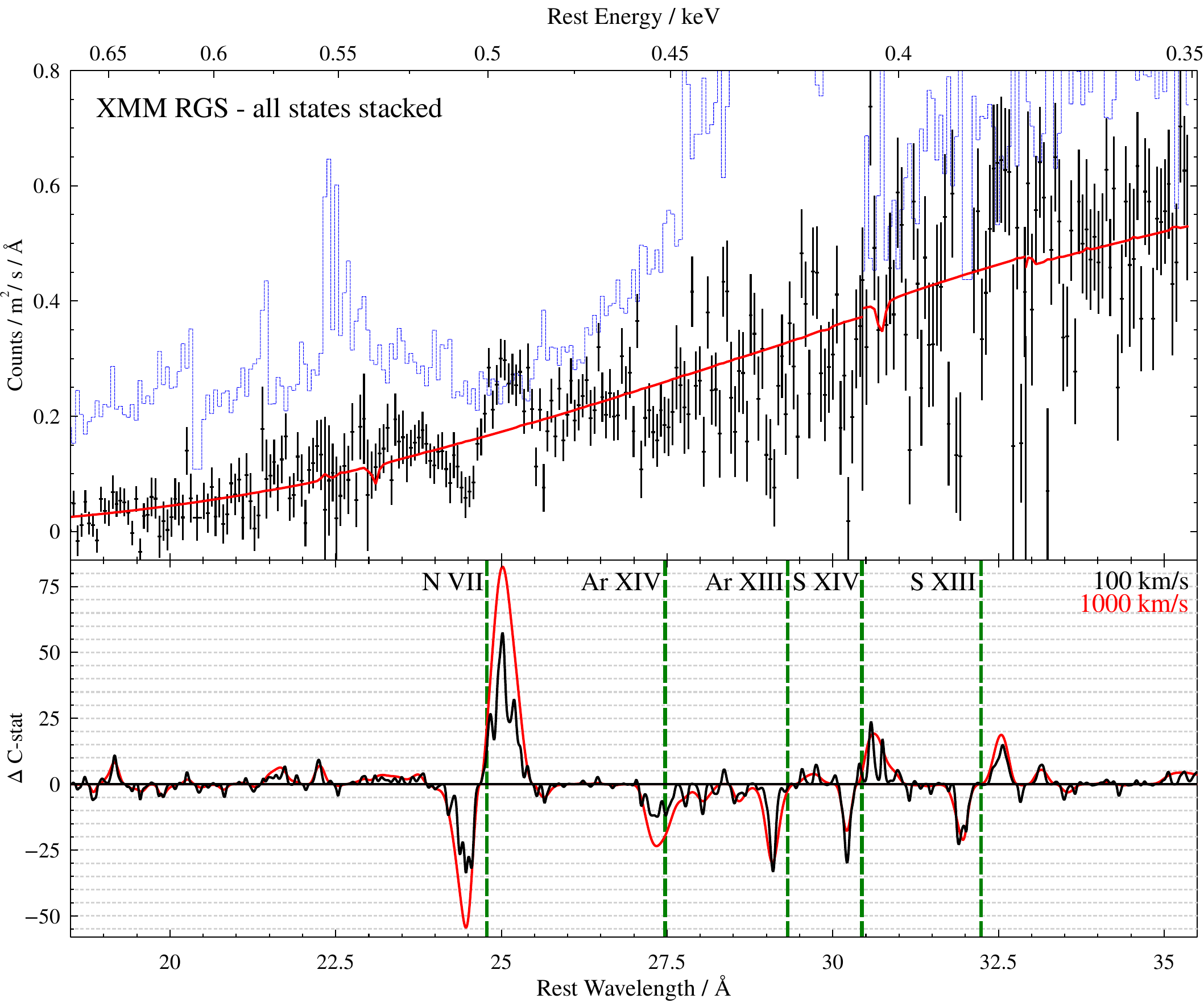}
\caption{Top panel: RGS grating spectrum of GSN 069, obtained by combining all available \xmm\ observations. Most of the wavelength range contains RGS 1+2 data, stacked and overbinned for visual purposes only, except the range between 30.5 and 32.9 \AA\ (in GSN 069 rest-frame) which only contains RGS 2 data. The best-fitting continuum model is shown in red and the observational background is in blue. Bottom panel: Gaussian line scan of the same spectrum. The Y axis defines the \delcstat\ fit improvement upon adding a line to the continuum-only model (multiplied by the sign of the spectral line normalization), with a line width of either 100 km/s (black) or 1000 km/s (red). Green dashed lines indicate the rest-frame wavelengths of transitions which can plausibly explain the strongest residuals.  \label{RGS_Gaussian_scan}}
\end{center}
\end{figure*}

The results of the Gaussian line scan for the stacked RGS spectrum are shown in Fig. \ref{RGS_Gaussian_scan}. The spectrum reveals two strong spectral lines exceeding \delcstat\ of 50 at around 25 \AA, as well as a large number of weaker (but still likely real) features with \delcstat\ between 15 and 30. In this figure we also list potential rest-frame line identifications of these features taken from the spectral fits of \citet{Miller+15} and \citet{Kosec+23b}, who studied the RGS spectra of tidal disruption events with similar spectral shapes as that of GSN 069. The strongest two features seen in GSN 069 are both close to the rest-frame wavelength of N VII, and appear to form a P-Cygni shaped profile around this transition. Given the \delcstat\ fit improvements of more than 50 for both the absorption and emission components, its statistical detection significance is thus far more than $4\sigma$. Most of the other residuals appear to cluster close to the rest-frame transitions of Ar XIII, XIV and S XIII, XIV. All of these transitions produced prominent absorption lines in the spectra of TDEs ASASSN-14li \citep{Miller+15} and ASASSN-20qc \citep{Kosec+23b}, but the Ar and S transitions are not usually strong in AGN warm absorber spectra as those outflows are typically less ionized \citep[e.g. Fig. 9 of][]{Kaastra+11}. We further note that all absorption residuals appear to the blue of the rest-frame wavelengths (with a visually consistent blueshift), while the emission residuals 
peak to the red of the rest-frame wavelengths, supporting a possible P-Cygni shape origin. Based on these spectral residuals and their positions, we conclude that GSN 069 likely exhibits an ionized outflow, observed both in absorption and emission.

To begin understanding the kinematic properties of this outflow, we use Gaussian lines to fit the two strongest residuals located near N VII (rest-frame wavelength of 24.77 \AA). The best-fitting absorption component wavelength is $24.48_{-0.03}^{+0.05}$ \AA, indicating a blueshift of $3500_{-600}^{+400}$ km/s if associated with N VII, and a FWHM width of $3800^{+1400}_{-1000}$ km/s. Conversely, the best-fitting emission component wavelength is $25.00_{-0.08}^{+0.04}$ \AA, resulting in a redshift of $2800_{-1000}^{+500} $ km/s, and has a FWHM width of $6000_{-900}^{+2400}$ km/s.

\subsection{State-Resolved Photoionization Analysis}

The presence of a large number of residuals in the RGS spectrum of GSN 069 motivates us to apply more complex photoionization 
spectral models which can describe multiple spectral features at once, and allow us to determine the physical properties of the absorbing or emitting plasma. To describe these features we primarily use the \textsc{pion} photoionization spectral model \citep{Mehdipour+16}. \textsc{pion} calculates transmission through a slab of plasma illuminated by the continuum currently loaded in \textsc{spex}. This comes at a significant computational cost, but ensures the most accurate calculation of the plasma ionization balance. \textsc{pion} can produce both absorption and re-emission from the illuminated plasma slab.

For this analysis, we also move away from a single stacked RGS spectrum and instead separately study the 3 RGS spectra of GSN 069 during its 3 different system states - Quiescence without QPEs, Quiescence with QPEs, and QPE state. The first two of these states offer sufficient quality RGS spectra for data analysis, which will inform us of any long-term trends in the evolution of the ionized outflow. The QPE dataset produces the lowest-quality RGS spectrum (due to the low exposure), but also the highest flux, which could be causing a bias in the overall RGS stack. Comparing the QPE and Quiescence with QPEs state spectra should allow us to probe how the plasma responds to the QPEs on the short timescales of ks and tens of ks.

\begin{deluxetable*}{c|ccccc|ccccc}
\tablecaption{Spectral fits of the ionized absorption and emission using the \textsc{pion} photoionization model, without coupling any parameters between the absorption and emission components. \label{PION_results_decoupled}}
\tablewidth{0pt}
\tablehead{
\colhead{State} & \multicolumn5c{Absorption component}  & \multicolumn5c{Emission component}  \\
\hline
\nocolhead{} & \colhead{$N_{\rm{H}}$} & \colhead{ $\log \xi$} & \colhead{Velocity}  & \colhead{Outflow}  & \colhead{$\Delta$} & \colhead{$N_{\rm{H}}$} & \colhead{$\log \xi$}  &  \colhead{Velocity}  & \colhead{Outflow}  & \colhead{$\Delta$} \\
\nocolhead{name} & \colhead{} & \colhead{} & \colhead{width}  & \colhead{velocity}  & \colhead{C-stat} & \colhead{} & \colhead{} &  \colhead{width}  & \colhead{velocity}  & \colhead{C-stat} \\
\colhead{} & \colhead{$10^{22}$ \pcm} & \colhead{erg cm s$^{-1}$} & \colhead{km/s}  & \colhead{km/s}  & \colhead{} & \colhead{$10^{22}$ \pcm} & \colhead{erg cm s$^{-1}$} & \colhead{km/s}  & \colhead{km/s}  & \colhead{}
}
\startdata 
Quiescent& \multirow{2}{*}{$1.1_{-0.2}^{+0.3}$}&\multirow{2}{*}{$3.86_{-0.12}^{+0.11}$}&\multirow{2}{*}{$1100 \pm 200$}&\multirow{2}{*}{$-2900 \pm 300$}&\multirow{2}{*}{53.33}&\multirow{2}{*}{$0.34_{-0.10}^{+0.11}$}&\multirow{2}{*}{$3.0 \pm 0.2$}&\multirow{2}{*}{$2700_{-1000}^{+800}$}&\multirow{2}{*}{$100_{-900}^{+1000}$}&\multirow{2}{*}{23.76}\\
without QPEs&&&&&&&&&&\\
\hline
Quiescent&\multirow{2}{*}{$1.7 \pm 0.4$}&\multirow{2}{*}{$4.57 \pm 0.13$}&\multirow{2}{*}{$490_{-80}^{+90}$}&\multirow{2}{*}{$-2340 \pm 120$}&\multirow{2}{*}{77.31}&\multirow{2}{*}{$1.3 \pm 0.4$}&\multirow{2}{*}{$3.95_{-0.22}^{+0.20}$}&\multirow{2}{*}{$1500_{-400}^{+500}$}&\multirow{2}{*}{$2700_{-300}^{+400}$}&\multirow{2}{*}{46.37}\\
with QPEs&&&&&&&&&&\\
\hline
\multirow{2}{*}{QPE} &\multirow{2}{*}{$1.0_{-0.4}^{+0.7}$}&\multirow{2}{*}{$4.6 \pm 0.2$}&\multirow{2}{*}{$400_{-200}^{+400}$}&\multirow{2}{*}{$-1700_{-400}^{+300}$}&\multirow{2}{*}{20.12}&\multirow{2}{*}{ $0.02_{-0.02}^{+0.09}$}&\multirow{2}{*}{3.95*}&\multirow{2}{*}{1500*}&\multirow{2}{*}{2700*}&\multirow{2}{*}{0.04}\\
&&&&&&&&&&\\
\enddata
\tablecomments{
$^{*}$Parameter value is fixed to the best-fitting value from the spectral fit of Quiescence with QPEs.}
\end{deluxetable*}

\subsubsection{Quiescent states with and without QPEs}

The spectrum of Quiescence without QPEs is the highest signal-to-noise dataset since GSN 069 generally shows higher time-averaged X-ray flux (by about a factor of two) in that state compared with Quiescence with QPEs \citep{Miniutti+23a}. Nevertheless, the two quiescence state spectra show comparable spectral fitting results. The baseline continuum is described using a \textsc{dbb} component with a best-fitting temperature of $0.122 \pm 0.002$ keV for Quiescence without QPEs and $0.109 \pm 0.002$ keV for Quiescence with QPEs. This is absorbed by Galactic, nearly neutral \textsc{hot} absorption, same as was done in the Gaussian line scan and we also use the \textsc{reds} model to take into account the redshift of GSN 069. Additionally, we tested for the presence of neutral absorption in the host galaxy of GSN 069 by adding an extra \textsc{hot} component redshifted by $z=0.018$, but did not detect any significant column density.

\begin{figure*}
\begin{center}
\includegraphics[width=\textwidth]{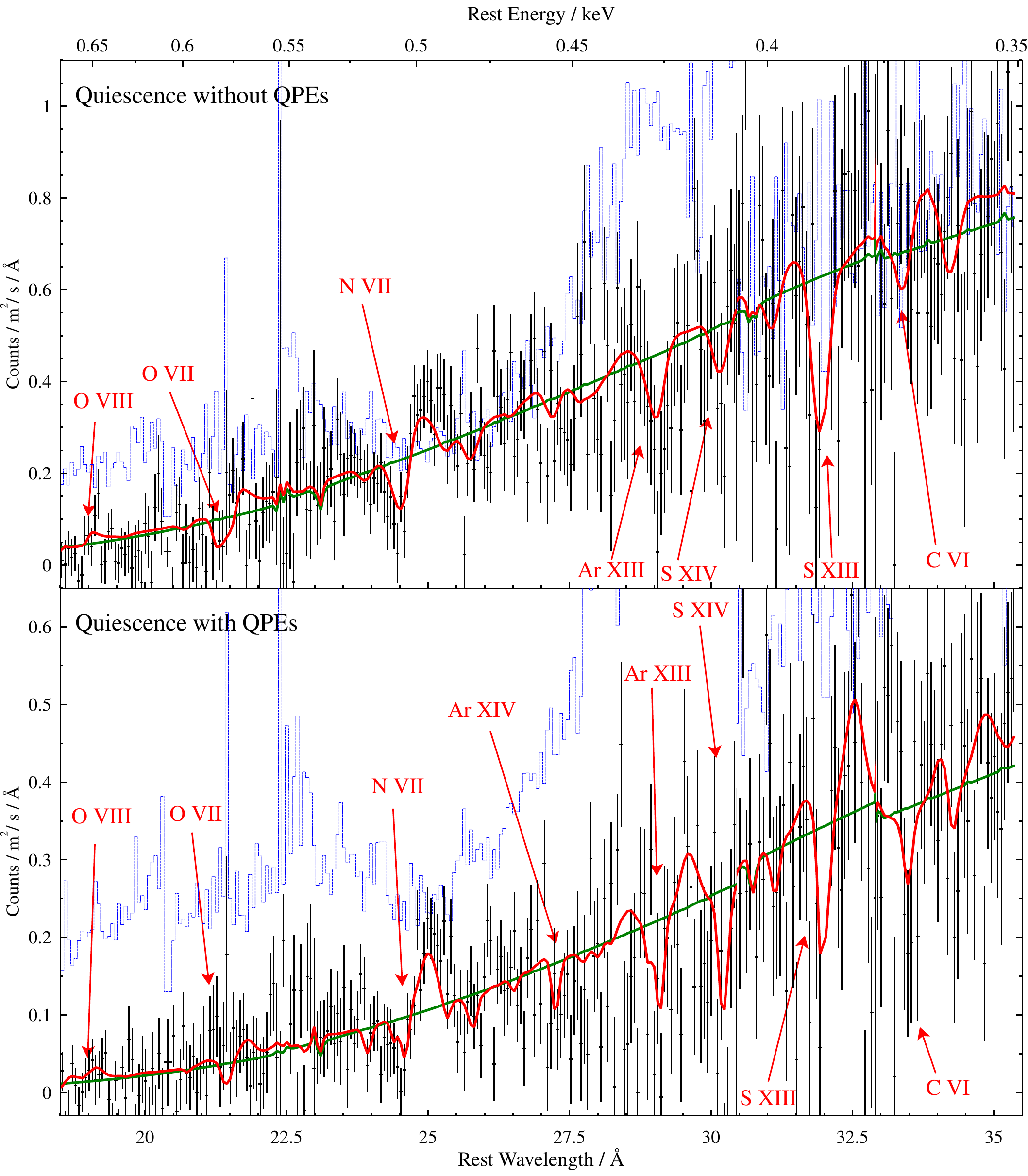}
\caption{RGS spectrum of GSN 069 during Quiescence without QPEs (top panel) and Quiescence with QPEs (bottom panel). Most of the wavelength range (shown in GSN 069 rest-frame) contains RGS 1+2 data, stacked and overbinned for visual purposes only, except the range between 30.5 and 32.9 \AA\ which only contains RGS 2 data. The best-fitting continuum-only model is shown in green, while the outflow model is in red, and contains both ionized absorption and emission (the decoupled spectral model). The most prominent spectral features are identified with red labels. The background spectrum is in blue.  \label{Spectrum_noQPE}}
\end{center}
\end{figure*}

We then apply two \textsc{pion} components to describe the ionized absorption and emission separately, with the expectation (based on the Gaussian scan results, Fig. \ref{RGS_Gaussian_scan}) that the absorption will be blueshifted, while the emission will be redshifted. Here, the absorber is assumed to be fully covering the X-ray emission region from our point of view (\textsc{pion} parameter \textsc{fcov} equal to 1) and the emitter is assumed to cover the full $4\pi$ solid angle from the point of view of GSN 069 (\textsc{pion} parameter $\Omega$ equal to 1). As a first step, we fit the absorption and emission components completely independently (without coupling any spectral parameters). Therefore, this fit does not assume any physical connection between the absorption and emission lines. We fit for the column density, ionization parameter \logxi, systematic (outflow) velocity and the velocity width of both components. We also recover the fit improvement \delcstat\ upon adding the ionized absorber over the baseline (\textsc{dbb}) continuum, and the fit improvement upon adding the ionized emitter to the baseline containing continuum and ionized absorption. The results of this spectral fit for both the quiescent states are listed in Table \ref{PION_results_decoupled}, and shown in Fig. \ref{Spectrum_noQPE}.

In the Quiescence without QPEs, we find that the absorber is moving at a velocity of $-2900 \pm 300$ km/s, while the emitter appears roughly stationary at $100^{+1000}_{-900}$ km/s (but with significant uncertainties). Both components are relatively highly ionized at \logxi$\sim3.86_{-0.12}^{+0.11}$ and $3.0 \pm 0.2$ for the absorption and emission, respectively, and show comparable column densities of $10^{21-22}$ \pcm. Importantly, the addition of the absorber to the baseline continuum fit is highly statistically significant as it improves the statistic by \delcstat$=53$. The emitter improves the spectral fit by somewhat less, \delcstat$\sim23$. This is still significant at about $3\sigma$ \citep{Pinto+20, Kosec+20b}. 

In the Quiescence with QPEs, the addition of ionized absorption and emission is highly significant over the baseline continuum with a fit improvement of \delcstat\ $\sim77$ and \delcstat\ $\sim46$, respectively. Both components show a column density of about $10^{22}$ \pcm\ and high ionization parameters \logxi\ of $4.57 \pm 0.13$ (absorber) and $3.95^{+0.20}_{-0.22}$ (emitter). The absorber is moving with a velocity of $-2340 \pm 120$ km/s, while the emitter is redshifted by $2700^{+400}_{-300}$ km/s.

The strongest spectral feature of both the absorber and the emitter is the N VII transition, where our signal-to-noise is highest, but other notable transitions that are present include O VII, Ar XIII, S XIV, S XIII and C VI. We note that while during the Quiescence without QPEs the absorption appears to be stronger than the emission, the opposite is true for the Quiescence with QPEs where the ionized emission dominates, particularly in the N VII transition.

The best-fitting $0.3-2.0$ keV X-ray flux is $(6.7 \pm 0.2) \times 10^{-13}$ erg s$^{-1}$ cm$^{-2}$ for Quiescence without QPEs and $(3.55 \pm 0.11) \times 10^{-13}$ erg s$^{-1}$ cm$^{-2}$ for Quiescence with QPEs.

The relative similarity of the column density and the ionization parameter of the absorber and the emitter motivates us to couple these components. We couple both the column density and the ionization parameter \logxi\, and instead introduce a new variable parameter, $\Omega$, the solid angle of the ionized emitter as seen from the ionizing source. Physically, this combination would represent a 3D outflow launched from GSN 069 in a range of directions (but not necessarily covering $4\pi$ of the sky of GSN 069). The outflow is seen directly along the line of sight to GSN 069 in absorption, as well as through re-emission from the outflow regions which are out of our line of sight towards GSN 069. The remaining parameters of this more constrained spectral model, the velocity width and outflow velocity, are left decoupled.

We caution, however, that this is not a fully accurate `P-Cygni' 3D model. In our first-order model, the kinematics of the emitting gas are given phenomenologically with the velocity width (derived from line widths) rather than using the ratio of the ionized absorption and emission velocities based on the geometric projection of the 3D outflow properties. More complex spectral models such as \textsc{wine} \citep{Luminari+18} or \textsc{xrade} \citep{Matzeu+22} would likely provide a more physical 3D kinematic description of the outflow. The application of more advanced wind spectral models is beyond the scope of this first discovery paper.

The results of this second spectral fit are shown in Table \ref{PION_results_coupled}. The coupled absorption-emission fit is slightly worse than the previous decoupled one (as it has fewer free parameters), and the addition of the coupled emission improves the absorption-only fit of the Quiescence without QPEs by \delcstat $\sim17$, and of the Quiescence with QPEs by \delcstat $\sim38$. Importantly, we find a solid angle $\Omega/4\pi$ of $0.9 \pm 0.3$ for Quiescence without QPEs and $1.1_{-0.3}^{+0.4}$ for Quiescence with QPEs. In other words, with such a best-fitting solid angle, the emitter is consistent with covering the whole $4\pi$ sky from the point of view of GSN 069. We note that this statement about the outflow solid angle is made purely based on the strength of the emission lines in the \textsc{pion} model, and not based on any line kinematics (or a more physical P-Cygni line shape analysis).

We also note that the outflow velocity of the emitter in the Quiescence without QPEs shifts to $2900_{-800}^{+900}$ km/s (which may explain the observed difference in the fit quality) and is now consistent with the redshift of the emitter during Quiescence with QPEs. This likely occurs due to the limited signal-to-noise in the emission component.


\begin{deluxetable*}{c|cccc|cccc}
\tablecaption{Spectral fits of the ionized absorption and emission using the \textsc{pion} photoionization model, while coupling the column density and ionization parameter between the absorption and emission models.\label{PION_results_coupled}}
\tablewidth{0pt}
\tablehead{
\colhead{State} & \multicolumn4c{Absorption component}  & \multicolumn4c{Emission component}  \\
\nocolhead{} & \colhead{$N_{\rm{H}}$} & \colhead{ $\log \xi$} & \colhead{Velocity}  & \colhead{Outflow}   & \colhead{Solid}  &  \colhead{Velocity}  & \colhead{Outflow}  & \colhead{$\Delta$} \\
\nocolhead{name} & \colhead{} & \colhead{} & \colhead{width}  & \colhead{velocity}  & \colhead{angle} &  \colhead{width}  & \colhead{velocity}  & \colhead{C-stat} \\
\colhead{} & \colhead{$10^{22}$ \pcm} & \colhead{erg cm s$^{-1}$} & \colhead{km/s}  & \colhead{km/s}  & \colhead{$\Omega/4\pi$} & \colhead{km/s}  & \colhead{km/s}  & \colhead{}
}
\startdata 
Quiescent & \multirow{2}{*}{$0.9 \pm 0.2$}&\multirow{2}{*}{$4.13_{-0.10}^{+0.09}$}&\multirow{2}{*}{$600_{-200}^{+400}$}&\multirow{2}{*}{$-2700^{+800}_{-300}$}&\multirow{2}{*}{$0.9 \pm 0.3$}&\multirow{2}{*}{ $2200_{-500}^{+600}$}&\multirow{2}{*}{$2900_{-800}^{+900}$}&\multirow{2}{*}{16.99}\\
without QPEs&&&&&&&&\\
\hline
Quiescent & \multirow{2}{*}{$1.4_{-0.3}^{+0.4}$}&\multirow{2}{*}{$4.51 \pm 0.12$}&\multirow{2}{*}{$470_{-70}^{+90}$}&\multirow{2}{*}{$-2340 \pm 120$}&\multirow{2}{*}{$1.1_{-0.3}^{+0.4}$}&\multirow{2}{*}{$1100 \pm 400$}&\multirow{2}{*}{$2800 \pm 300$}&\multirow{2}{*}{38.17}\\
with QPEs&&&&&&&&\\
\hline
\multirow{2}{*}{QPE} &\multirow{2}{*}{$1.0_{-0.4}^{+0.8}$}&\multirow{2}{*}{$4.6 \pm 0.2$}&\multirow{2}{*}{$400_{-200}^{+400}$}&\multirow{2}{*}{$-1700_{-400}^{+300}$}&\multirow{2}{*}{$0.06_{-0.06}^{+0.22}$}&\multirow{2}{*}{1100*}&\multirow{2}{*}{2800*}&\multirow{2}{*}{0.05}\\
&&&&&&&&\\
\enddata
\tablecomments{
$^{*}$Parameter value is fixed to the best-fitting value from the spectral fit of Quiescence with QPEs.}
\end{deluxetable*}

As a further step, we consider that the outflow has non-Solar elemental abundances. Recent work by \citet{Miller+23} on the TDE ASASSN-14li indicated a very high nitrogen over-abundance (N$~>100$ over Solar ratios) compared with other elements. This could also be the case in GSN 069. In fact, the UV spectra of GSN 069 (from the \textit{Hubble Space Telescope}) show over-abundance of nitrogen \citep{Sheng+21}.

We free the N abundance in the \textsc{pion} model for both the ionized absorber and the emitter, linking this parameter between the two components, and re-fit. For Quiescence without QPEs, the best-fitting N abundance is $ 11_{-5}^{+9}$ for the decoupled fit and $11_{-5}^{+8}$ for the coupled fit. The introduction of N abundance as a free parameter further improves the spectral fit by \delcstat$\sim17$ for the decoupled fit and by \delcstat$\sim24$ for the coupled fit. For the Quiescence with QPEs, we obtain an N abundance of $ 40_{ -18}^{+43}$ for the decoupled spectral fit and an N abundance of $50_{-20}^{+40}$ for the coupled fit. Freeing the N abundance improves the fit by \delcstat$\sim40$ (for the decoupled fit) or by \delcstat$\sim32$ (for the coupled fit). Therefore, it appears that N is likely strongly overabundant in comparison with other metals in GSN 069. As we have limited data quality, we only perform this procedure for N, which shows the strongest spectral feature (N VII transition). Caution is still required when interpreting this result considering we have a limited number of plasma line detections to make this measurement, but it is statistically significant and in line with the UV spectroscopic results by \citet{Sheng+21}.

As a final step, we perform a combined fit of both quiescence states to determine the most likely N over-abundance value. We use the coupled absorption-emission model and simultaneously fit the two RGS datasets while keeping all of the physical parameters decoupled except the N abundance. This results in a best-fitting N abundance of $24_{-10}^{+17}$. Fitting for the N abundance slightly shifts the other best-fitting ionized absorber and emitter parameters, but for nearly all parameters these shifts are within $1-2\sigma$ uncertainties of the original values.

\subsubsection{QPE state}

The QPE RGS spectrum has the highest count rate, but also the briefest total exposure, and thus the lowest total count statistics. During the QPEs, GSN 069 is known to harden, a spectral shape which can be reproduced by the quiescent state \textsc{dbb} component plus an extra soft blackbody with a temperature of about 100 eV. We therefore add this blackbody to the baseline continuum from the Quiescence with QPEs, and fit for its temperature and normalization. To reduce the fitting degeneracy, we fix the temperature of the \textsc{dbb} component to the best-fitting value from the Quiescence with QPEs. 

\begin{figure*}
\begin{center}
\includegraphics[width=\textwidth]{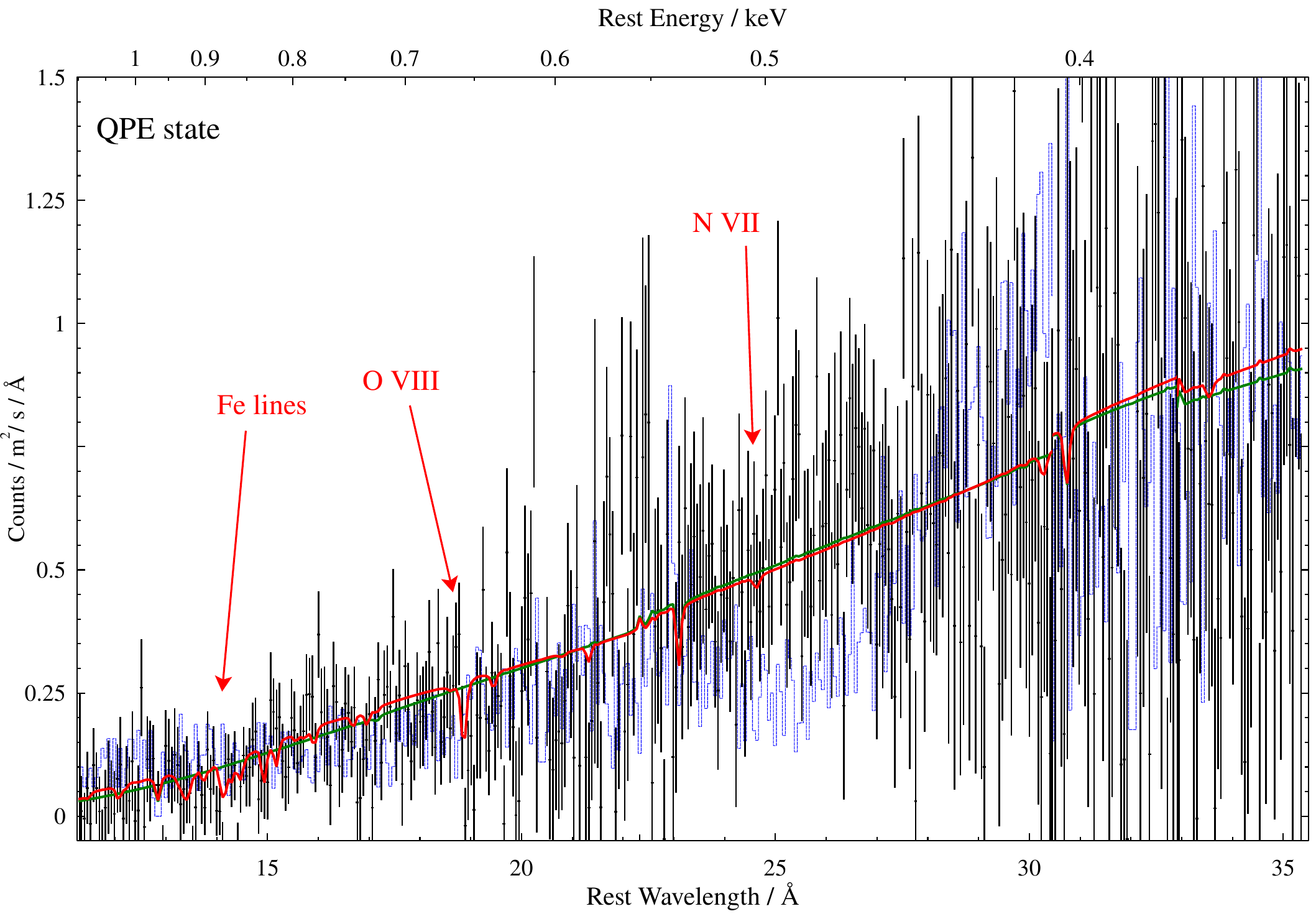}
\caption{Same as Fig. \ref{Spectrum_noQPE}, but showing the QPE spectral state. \label{Spectrum_flare}}
\end{center}
\end{figure*}

We then follow the same procedure as for the other two states. The results are shown in Tables \ref{PION_results_decoupled} and \ref{PION_results_coupled}, and in Fig. \ref{Spectrum_flare}. We find evidence for blueshifted absorption even in this lower quality spectrum. Including the absorber improves the fit by \delcstat$\sim20$, so it is detected with a significance of about $3\sigma$. However, the detection likelihood is boosted by the fact that the absorber properties during the QPE state are consistent with those during Quiescence with QPEs. We do detect a potential shift in the outflow velocity from $-2340 \pm 120$ km/s during that state compared with $-1700^{+300}_{-400}$ km/s during the QPE state. With the harder SED during the QPE state, the N VII transition is no longer the strongest spectral feature. Instead, we observe absorption from O VIII, and from the Fe transitions in the Fe unresolved transition array (UTA).

Surprisingly, we do not detect any ionized emission in the QPE state. The addition of an emission \textsc{pion} component does not significantly improve the spectral fit. We calculate the upper limit on the emitter column density or its solid angle by assuming the same ionization parameter, velocity width and outflow velocity as during the Quiescence with QPEs. We obtain a column density of less than $0.11 \times 10^{22}$ \pcm\ and a solid angle $\Omega/4\pi$ of less than $0.28$. 

We also tested for non-Solar N abundance during the QPE state, but due to the lower data quality, we were not able to obtain any meaningful constraints using this dataset. The best-fitting $0.3-2.0$ keV X-ray flux for this state is $(1.13 \pm 0.07) \times 10^{-12}$ erg s$^{-1}$ cm$^{-2}$.

\subsection{Further data quality checks}
\label{sec:dataqualitychecks}

For the RGS spectra of each of the 3 states, we also consider other interpretations for the observed spectral features. First we consider that some of these features could originate from absorption in the hot phase of the Galactic interstellar medium (ISM). To test this possibility, we add an extra \textsc{hot} absorption component in our baseline continuum (disk blackbody) fits. To model hot ISM, we fix the temperature of this component (at rest with z=0) to 0.1 keV. Then the spectral fit is re-run, and we determine the fit improvement \delcstat\ as well as the best-fitting hot ISM column density. We found that the addition of this component is not statistically significant in any of the 3 state spectra, with fit improvements of at most \delcstat$\sim1$. The best-fitting column density was less than $10^{20}$ \pcm\ in all cases. Hot ISM can therefore not explain the observed absorption features.

Additionally, as the GSN 069 X-ray flux is very low, the source spectrum is on the same level or below the RGS background flux level for a significant fraction of the wavelength range. This is especially important for the spectrum of Quiescence with QPEs, which is background-dominated throughout all of the wavelength range. In general, RGS background is relatively stable and so the background subtraction should be accurate. Nevertheless, we perform an additional check of our spectra with alternative background spectra. For this check, instead of instrumental background accumulated during the actual observations (from the edges of the RGS CCD chips), we use blank field RGS backgrounds. In this case the background subtraction is significantly worse than the instrumental background subtraction. Across the wavelength range, we observe broad continuum residuals correlated with background flux, and the agreement between RGS 1 and RGS 2 data is no longer within 10\% as it is with instrumental background. Nevertheless, the same spectral features are still present in background-subtracted source spectra. We find that the addition of an ionized absorber improves the Quiescent state without QPEs baseline (\textsc{dbb}) continuum fit by \delcstat$=30$, and the Quiescent state with QPEs baseline fit by \delcstat$=82.13$. As the source flux is significantly higher during the QPE state, we did not perform this check for the QPE spectrum.

Thirdly, we consider that we may not have matched the observed spectral residuals with the correct elemental transitions. To verify our identification, we perform a systematic, automated grid search of the RGS spectra for plasma in absorption over a broad range of possible ionization parameters, outflow velocities and velocity widths. This search is described in further detail in Appendix \ref{app:systematic_search}, and shows that we have indeed found the global best fit for the ionized absorber physical properties in the GSN 069 RGS data.

Finally, we test alternative ionizing mechanisms for the observed features. With the application of the \textsc{pion} model, we explicitly assumed that the plasma is photoionized, which may not necessarily be the case. Instead, the plasma may be collisionally ionized, for example by shocks, as was recently found for one of the ionization zones of the outflow in the AGN NGC 4051 \citep{Ogorzalek+22}. We test this scenario with collisional ionization spectral models. We use the \textsc{hot} model to describe the absorption features and the \textsc{cie} model \citep{Kaastra+96} to describe the emission. We found that \textsc{cie} at Solar abundances cannot reproduce the observed emission as the strongest observed emission line is N VII. A high abundance of N compared with O (N/O$\sim 40-70$), and a high temperature ($\sim0.3$ keV) is required to observe a strong N VII line without observing a strong N VI line and not detecting strong O VIII or O VII emission lines. However, even then the collisional spectral models are not preferred over the photoionization models presented above. The photoionization models are preferred in both quiescent state spectral fits by at least \delcstat$\sim15$. Finally, we test an alternative where the emission lines are produced by photoionization (described with \textsc{pion}), but the absorption lines are from collisionally ionized plasma. However, this solution is not statistically preferred in any of the spectral fits (in any of the 3 GSN 069 states) over pure photoionization.

\section{Residuals in \xmm\ EPIC spectra}
 \label{sec:epicresults}

\citet{Miniutti+23a} previously showed that \xmm\ EPIC spectra of GSN 069 in quiescence also contain a strong absorption residual, which can be reproduced with ionized absorption. This EPIC residual is located at $0.6-0.8$ keV, where our RGS datasets have little signal during both quiescent states (the spectra lose all source counts above $\sim0.65$ keV, Fig. \ref{Spectrum_noQPE}). Hence, the strong lines that we observe in the RGS are different spectral features than this $0.6-0.8$ keV residual seen in EPIC data. For comparison, the N VII lines we resolve in the RGS data are located at $\sim0.5$ keV, and most of the remaining notable residuals are at even lower energies (Fig. \ref{RGS_Gaussian_scan}). By applying the \textsc{xstar} photoionization model, \citet{Miniutti+23a} were able to fit the EPIC residuals with a warm absorber with an ionization parameter \logxi\ of about 0.5 and a column density of $(5-7)\times10^{21}$ \pcm. However, this interpretation is only tentative given the limited spectral resolution of EPIC pn and MOS detectors of about 100 eV, resulting in a resolving power of $R<10$ in this energy range.

Now, the RGS data on GSN 069 can help identify the nature of these EPIC residuals. Any ionized plasma, for example a warm absorber fitted to the EPIC data, will also produce spectral features across the RGS energy band. Conversely, the ionized emission and absorption models we fitted to the RGS data will also produce spectral features across the EPIC energy band. Thus, by combining these two datasets, we will obtain a better understanding of the ionized plasma in the vicinity of GSN 069. We begin addressing this question here, but stress that we perform only a preliminary analysis of the EPIC dataset on GSN 069. A rigorous combined RGS-EPIC analysis is not straightforward for very soft X-ray sources such as GSN 069 \citep[as discussed in][]{Kosec+23b} and is beyond the scope of this discovery paper.

\subsection{Quiescent states with and without QPEs}

\begin{figure*}
\begin{center}
\includegraphics[width=\textwidth]{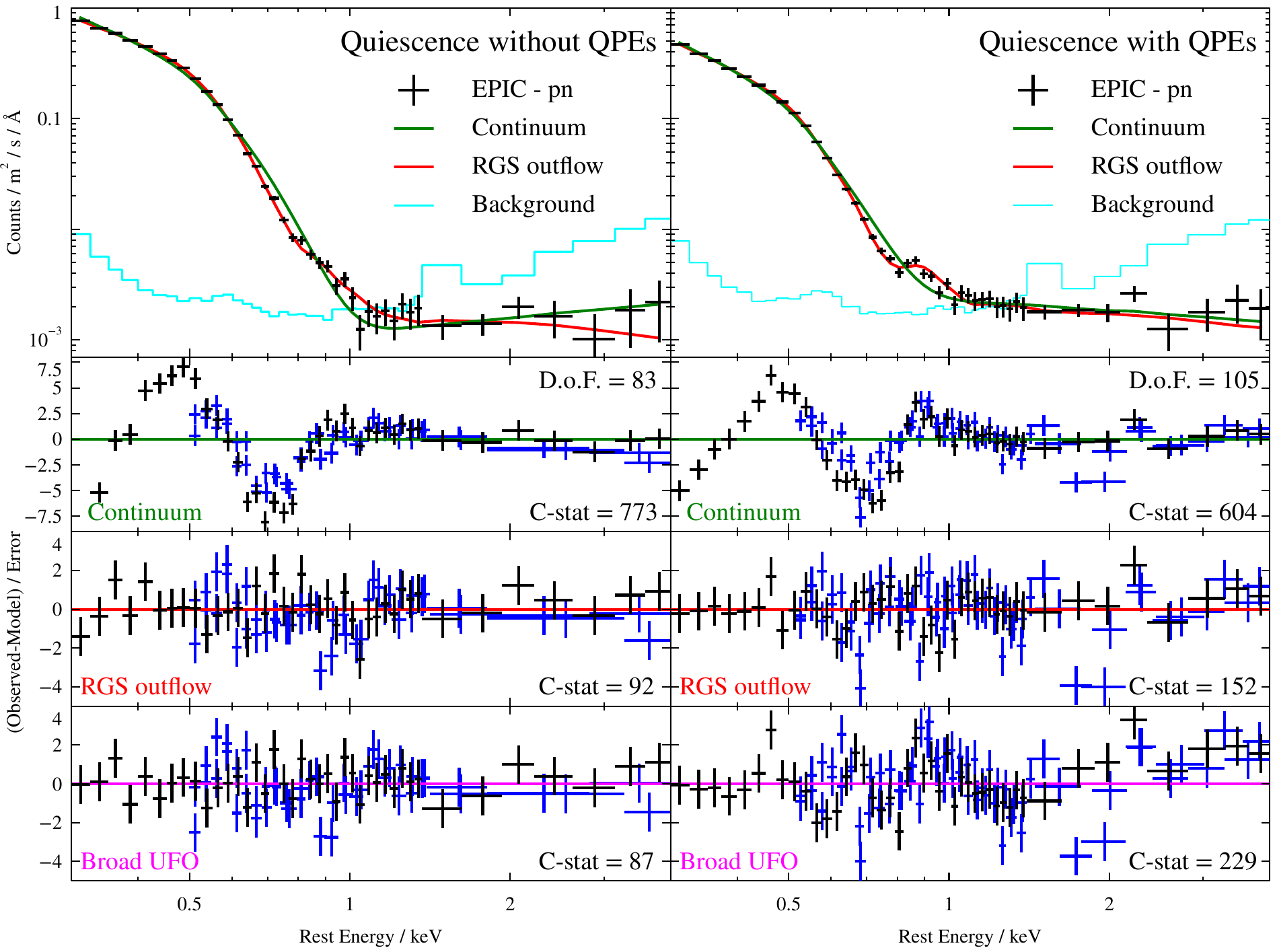}
\caption{\xmm\ EPIC spectrum of GSN 069 during Quiescence without QPEs (left) and Quiescence with QPEs (right). Top panels: EPIC-pn data (MOS 1 and 2 not shown for clarity) fitted with a base continuum model (green) and with the RGS outflow model (red). Background is shown in cyan. The broad UFO models appear very similar to the RGS outflow models (red) in this log-log plot and so are not shown for clarity. Lower three panels: residuals (pn in black, MOS 1 and 2 in blue) to the continuum, RGS outflow, and broad UFO model spectral fits, respectively. Values of the C-stat fitting statistic are listed for each fit on the bottom right. The number of degrees of freedom is given for the continuum fit on the top right. The RGS outflow and Broad UFO models both have 4 less D.o.F. compared with the base continuum. \label{EPIC_quiescence}}
\end{center}
\end{figure*}

\begin{figure}
\includegraphics[width=\columnwidth]{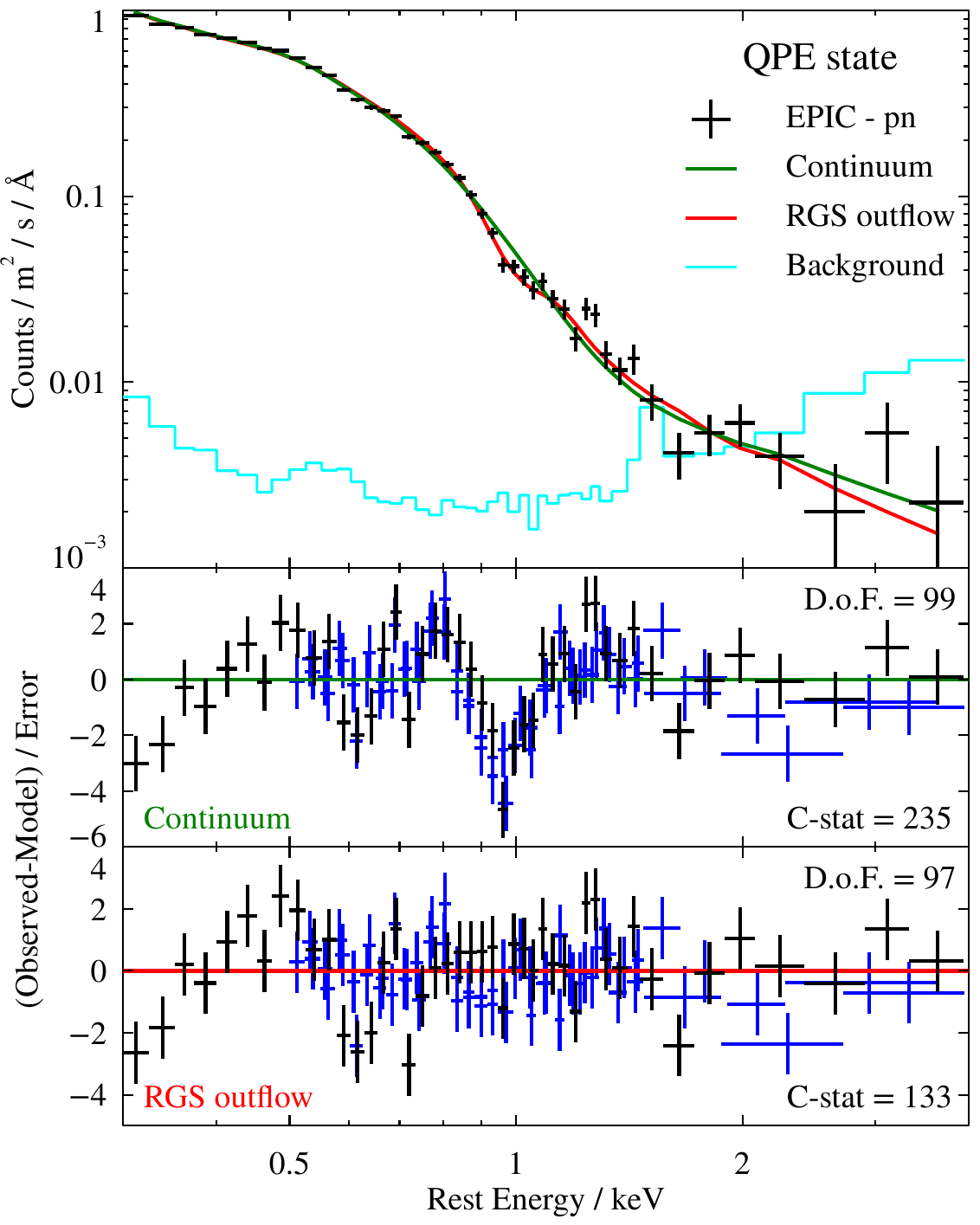}
\caption{\xmm\ EPIC spectrum of GSN 069 in the QPE state. Top panel: EPIC-pn data (MOS 1 and 2 not shown for clarity) fitted with a base continuum model (green) and with the RGS outflow model (red). Background is shown in cyan. Lower two panels: residuals (pn in black, MOS 1 and 2 in blue) to the continuum and RGS outflow model spectral fits, respectively. Values of the C-stat fitting statistic are listed for each fit on the bottom right. The number of degrees of freedom is given for each fit on the top right. \label{EPIC_flare}}
\end{figure}

As described in Section \ref{sec:data}, alongside the state-resolved RGS data reduction, we also perform EPIC (pn and MOS) data reduction, and hence obtain simultaneous EPIC spectra. To begin describing these data, we first fit them with a base continuum model. The spectra are analyzed in the energy range between 0.3 keV and 4.0 keV (EPIC-pn) and 0.5 keV to 4.0 keV (EPIC-MOS), and are shown in the top panels of Fig. \ref{EPIC_quiescence} (Quiescence with and without QPEs) and Fig. \ref{EPIC_flare} (QPE state). Above $1.0-1.5$ keV, the source is comparable or weaker than the background, but we do observe a hard tail inconsistent with a Wien tail of a blackbody or a disk blackbody. The presence and the nature of this hard component is speculative in GSN 069, but appears to be statistically significant in these stacked data. We will not focus on the physical interpretation of this new continuum component, but it must be included in our fitting to correctly model the absorption residuals. Above 4 keV, we observe no significant source signal, and so ignore all higher energy counts.

The continuum model is similar to the one we used for the RGS data, but with an additional powerlaw component describing this hard tail. The powerlaw has a typical slope in the range between 2.5 and 3.0 (depending on the exact spectral fit and GSN 069 state), softer than a standard AGN corona emission. To make this new continuum SED more physical, we add a lower exponential cutoff to prevent the powerlaw luminosity from diverging at low energies (and crashing the photoionization SED calculation), and set the cutoff parameter to be equal to the disk blackbody temperature (assuming that the accretion disk provides seed photons for this powerlaw emission). For this operation we use the \textsc{etau} model in \textsc{spex}. Therefore, the baseline continuum model is \textsc{hot~$\times$~reds~$\times$~(dbb+pow~$\times$~etau)} for both quiescent states, and \textsc{hot~$\times$~reds~$\times$~(dbb+bb+pow~$\times$~etau)} for the QPE state.

The results of these continuum spectral fits are shown in the top two panels of Fig. \ref{EPIC_quiescence} (Quiescence with and without QPEs) and Fig \ref{EPIC_flare} (QPE state). In both quiescent states, we observe a highly significant absorption residual at $0.6-0.8$ keV, in agreement with the results of \citet{Miniutti+23a}. The strength of this residual is clear from the high C-stat fit statistics of both Quiescence without QPEs (C-stat=773 for 83 D.o.F.) and Quiescence with QPEs (C-stat=604 for 105 D.o.F.) spectral fits. On the other hand, a dominant $0.6-0.8$ keV residual is not present in the QPE state spectrum, and instead we observe (a weaker) absorption residual at $0.9-1.0$ keV.

We begin the photoionization analysis by reproducing the warm absorber spectral fit from \citet{Miniutti+23a} for both quiescent states, using a \textsc{pion} component at rest (no outflow velocity). The best-fitting column density is around $3\times10^{21}$ \pcm, comparable with their result, however we obtain an ionization parameter \logxi\ of about -0.75. This discrepancy could be due to either differences between the \textsc{pion} and \textsc{xstar} codes \citep{Mehdipour+16}, or due to a different SED adopted between our work and \citet{Miniutti+23a}. In any case, such a warm absorber describes our EPIC residuals reasonably well, reducing the C-stat of the Quiescent state without QPEs spectral fit to only 132 (for 81 D.o.F) and the C-stat of the Quiescent state with QPEs fit to 222 (for 103 D.o.F.). 

However, this ionized absorption model produces strong narrow lines throughout the $0.3-1$ keV energy band, unresolved in EPIC data but resolved by the RGS. These lines are not present in the RGS data from the same GSN 069 states, and so warm absorption cannot be the explanation of the EPIC residuals. Using the RGS datasets, we place an upper limit of just $3\times10^{20}$ \pcm\ for a warm absorber at rest with \logxi\ equal to the best-fitting value from the EPIC analysis. This is a factor of 10 lower than what is required by the EPIC data.

The first alternative explanation for these EPIC residuals is that they are produced by the same ionized plasma which imprints the RGS spectral features. To investigate this possibility, we used the same \textsc{pion} models applied to RGS data to fit the EPIC spectra. This does not result in fit statistic improvement under the assumption of Solar abundances. However, the RGS data show intriguing evidence that the outflow abundances are non-Solar, with a strong N over-abundance. As a second step, we apply \textsc{pion} models modified to include the best-fitting N abundance of 24 (fixed) from the combined quiescent state RGS analysis. Given the limited spectral resolution of EPIC, we simplify the original \textsc{pion} photoionization models. We freeze all \textsc{pion} outflow velocities and velocity widths to the best-fitting values from the RGS analysis (coupled emission-absorption analysis with tied N abundance), and only fit for \textsc{pion} column densities and ionization parameters, in addition to the underlying continuum properties.

These models work much better with the EPIC data, and produce reasonable spectral fits, shown in the third row panels in Fig. \ref{EPIC_quiescence}. The statistic improves to C-stat=92 (79 D.o.F.) for Quiescence without QPEs, and to C-stat=152 (101 D.o.F.) for Quiescence with QPEs. Effectively, all of the systematic residuals are removed from the EPIC spectra. However, we note a discrepancy between the best-fitting outflow parameters from these fits and the RGS dataset analysis. The best-fitting column density for the ionized absorber (in both GSN 069 states) is $(3-4)\times10^{21}$ \pcm, and its ionization parameter \logxi\ is only $2.5-2.8$. This is much lower than found in the RGS analysis (Table \ref{PION_results_decoupled}). The best-fitting parameters of the ionized emitter roughly match those from the RGS analysis for the Quiescence with QPEs with a column density of $2.1_{-0.6}^{+0.9}\times10^{22}$ \pcm\ and \logxi\ of $4.82_{-0.09}^{+0.07}$, but no emitter is found in the EPIC spectrum of Quiescence without QPEs. Given the discrepancy in the ionization parameter of the absorber, we did not attempt a coupled analysis as with the RGS dataset.

This is a considerable discrepancy, but we note that some differences are expected due to the mismatch between the spectral resolution of the two instruments, due to the different energy bands adopted, as well as due to any residual calibration differences. Additionally, the ionizing SEDs are not the same in the RGS and EPIC analyses, especially in the harder band where we included a hard tail for the EPIC continuum model. To begin assessing the effect of these different SEDs, we added the hard tail component from the EPIC continuum model (with the slope and normalization fixed to the best-fitting values from EPIC data for each state) to the RGS analysis for each state. This did not significantly shift any of the best-fitting outflow properties in the RGS analysis. It is thus unlikely that SED differences are the main driver of the outflow parameter discrepancy. However, a more thorough combined RGS-EPIC analysis is required.

At this stage, it is not clear if the observed difference can be reconciled only with the instrumental and modeling reasons mentioned above. A possible alternative explanation is that the RGS outflow is multi-phase, and the limited signal-to-noise RGS data reveal just one phase as they are only sensitive below 0.65 keV, while another phase is more clearly imprinted on the higher-energy EPIC data, which is more sensitive above 0.6 keV (thanks to the resolving power improving with increasing energy). A detailed simultaneous RGS-EPIC analysis to investigate this hypothesis is beyond the scope of this discovery paper and will be performed elsewhere.

A second possibility is that the outflow seen in the RGS data is not the origin of the EPIC residuals, which are instead imprinted by a highly blueshifted, high-velocity width ultra-fast outflow (UFO). Such a broad UFO can produce a single dominant, highly broadened absorption feature and thus reproduce the observed residual in both quiescence spectra. This interpretation was proposed by \citet{Kara+18} and \citet{Pasham+24a} to fit similar broad absorption residuals observed between 0.5 and 1 keV in two TDEs. To test this hypothesis, we apply a UFO model to both quiescent state EPIC spectra, again using an extra \textsc{pion} photoionization component. The spectral fit quality is strongly improved for both states, down to C-stat=87 (79 D.o.F.) for Quiescence without QPEs and C-stat=229 (101 D.o.F.) for Quiescence with QPEs. Therefore, the spectral fit quality is comparable between the two interpretations for Quiescence without QPEs, while the first interpretation (RGS outflow) is somewhat better than the UFO hypothesis for Quiescence with QPEs. The best-fitting UFO properties are similar for both GSN 069 states. We obtain a column density in the range of $(4-10)\times10^{21}$ \pcm, an ionization parameter \logxi\ of $2.6-2.9$, a velocity width of $(0.13-0.17)$c (full width half-maximum), and a blueshift of $(0.23-0.30)$c. These parameters are comparable to the results of UFO modeling in TDEs \citep{Kara+18, Pasham+24a}.

Importantly, these UFO absorbers produce very broad absorption features, and thus imprint only weak features in the RGS spectra which have lower count rates and statistics. To assess if the UFO models are allowed by the RGS spectra, we added the best-fitting UFO absorbers to the previous continuum + ionized outflow spectral fits (of the RGS data) for both quiescent states. Doing this does not significantly modify these spectral fits apart from needing continuum normalization adjustment. After re-fitting these normalizations, the differences in fit quality are very small, in both cases lower than \delcstat$\sim6$. Therefore, the RGS data do not disagree with the UFO hypothesis.

\subsection{QPE State}

While we had difficulties reproducing the EPIC spectra of the quiescent states with the RGS outflow models, this is not the case for the QPE state EPIC spectrum. Instead of a strong $0.6-0.8$ keV residual, the QPE state spectrum shows a weaker feature at $0.9-1.0$ keV. An ionized absorber model with kinematics (velocity width and outflow velocity) fixed to those from the RGS spectral analysis, and an N overabundance of $N=24$ is a good fit to the EPIC data, improving the fit to C-stat=133 for 97 D.o.F. The fit improvement is clearly seen in the residual plots comparing this fit with the baseline continuum one (lower two panels in Fig. \ref{EPIC_flare}). The fit results in a best-fitting column density of $1.6^{+0.4}_{-0.3} \times 10^{22}$ \pcm\ and a \logxi\ of $5.03 \pm 0.10$. Both of these values are comparable within $2\sigma$ with the results for this component from the simultaneous RGS analysis. 

To see if a broad UFO could still be present in this spectrum, we added a \textsc{pion} component with the best-fitting `broad UFO' properties from Quiescence with QPEs. This significantly worsens the fit by \delcstat\ of about 100, and introduces an emission spectral residual around $0.7-0.8$ keV. In principle, the UFO could be over-ionized by the QPE flux and increase in \logxi. To test this, we free the UFO component ionization in the spectral fit. We are unable to constrain the component ionization parameter as its value hits the upper bound of the \textsc{pion} model, but we obtain a lower limit of \logxi\ $=6.2$ for the UFO component ionization. Hence, the ionization would have to increase by a factor of more than 1000 to agree with the EPIC QPE dataset, assuming its column density and kinematics do not vary over the course of individual QPEs. At such high ionization, practically all of the absorption signatures of this component are gone. We thus conclude that the EPIC QPE spectrum of GSN 069 can be entirely explained by the outflow detected in the RGS dataset and no UFO component is required in this state.

\section{Discussion} \label{sec:discussion}

We perform the first high-spectral resolution X-ray study of a quasi-periodically erupting source by leveraging the extensive archive of \xmm\ observations on GSN 069. In the stacked RGS spectrum (totalling almost 2 Ms of raw exposure time), we resolve an array of emission and absorption lines. The statistical detection significance of the strongest individual line, identified as N VII, by far exceeds $4\sigma$. Next we perform a dataset split based on the GSN 069 flux state and apply photoionization spectral models. We confirm that an outflow is significantly detected both in absorption and emission in Quiescence without QPEs (when no QPE activity is observed) and Quiescence with QPEs (in-between individual QPEs), and detected only in absorption at around $3\sigma$ significance in the QPE state (during the QPEs). After establishing that the observed spectral residuals cannot be due to Poisson noise, we performed a number of checks to make sure we interpret their nature correctly. These features cannot be consistently explained by an ionized absorber with a different ionization state, outflow velocity, or velocity width (Appendix \ref{app:systematic_search}). The features are observed even if we use blank field backgrounds instead of observational background files. Finally, they cannot originate from hot ISM in our Galaxy.

To sum up, we detect an ionized outflow in GSN 069, seen both in absorption and emission. The column density of the outflow is about $10^{22}$ \pcm, with an ionization parameter \logxi\ of $3.9-4.6$ and a projected velocity of $1700-2900$ km/s. The observed absorption and emission components can be linked together, and interpreted as a single outflow moving away from GSN 069 with a velocity of about 2500 km/s, consistent with covering the full $4\pi$ sky from the point of view of the ionizing source.

This is the first confirmed detection and detailed characterization of an ionized outflow in a QPE system. Such phenomena may be present in other sources of this class, but it is very challenging to detect them. By exploiting nearly 2 Ms of \xmm\ observations, we pushed the RGS instrument to the limit and detected an ionized outflow in this relatively faint X-ray source. Megasecond-duration observational campaigns (with high-resolution X-ray instruments), or next-generation X-ray instruments will be necessary to detect outflow signatures in the fainter QPE sources \citep[such as RX J1301.9+2747,][]{Giustini+20} at a similar level as we did here in GSN 069.

\begin{figure}
\includegraphics[width=\columnwidth]{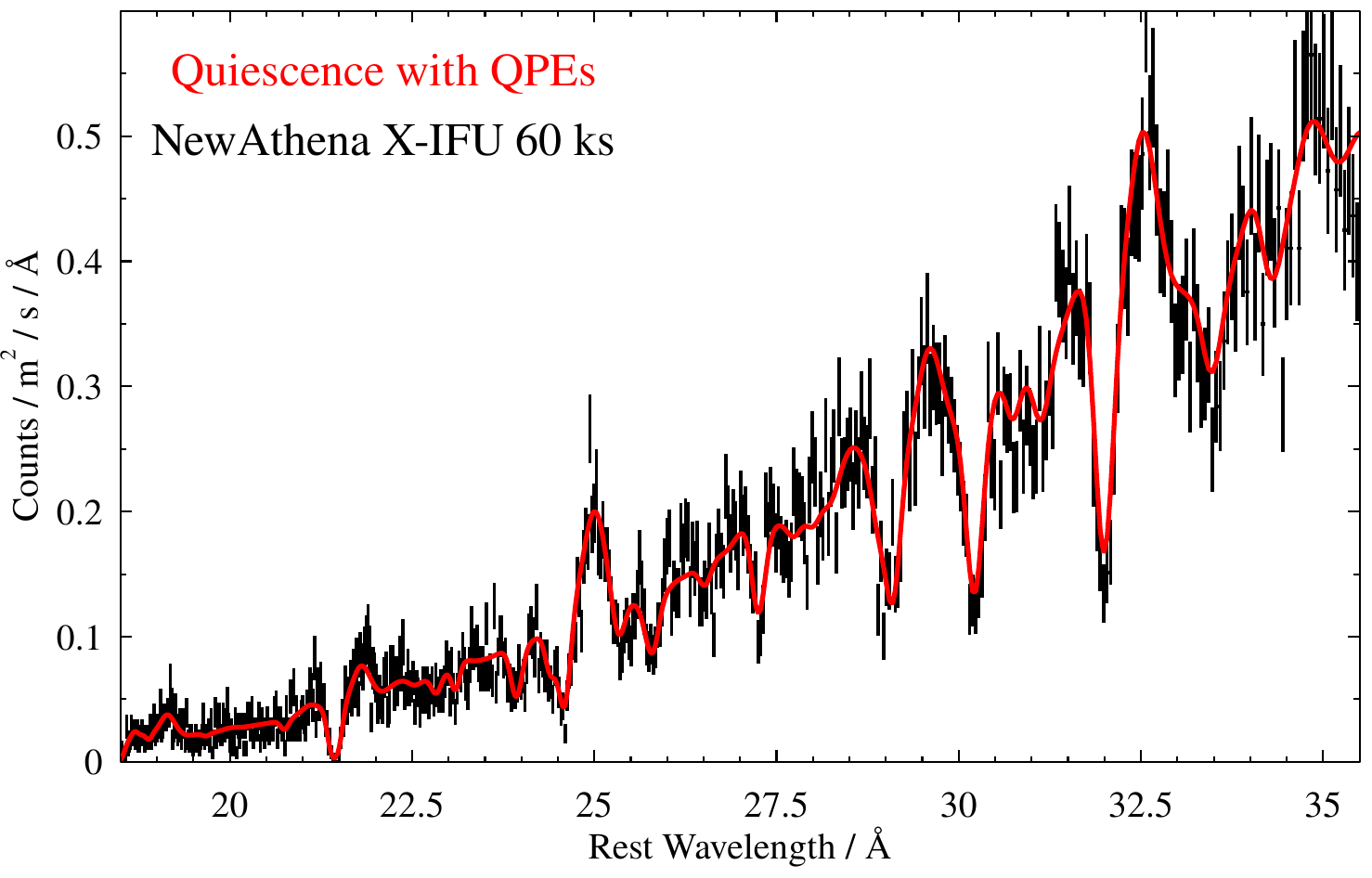}
\caption{Simulation of a 60 ks \athena\ X-IFU observation of GSN 069 in Quiescence without QPEs. We used the decoupled absorption-emission spectral model, the same as shown in Fig. \ref{Spectrum_noQPE} (bottom panel). \label{NewAthena_sim}}
\end{figure}

Future instruments offering high-spectral resolution in the soft X-rays as well as improved effective area, such as \athena\ \citep{Nandra+13}, \textit{Line Emission Mapper} \citep{Kraft+22} or \textit{Arcus} \citep{Smith+20}, are going to be crucial in revealing further properties of ionized outflows in GSN 069 and other QPE systems. To judge the performance of one of these instruments for the case of GSN 069, we performed a 60 ks \athena\ X-IFU \citep{Barret+23} observation simulation of the Quiescent state with QPEs (the faintest state of GSN 069, for a conservative simulation). We simulated the best-fitting decoupled absorption-emission outflow model and used the newest \athena\ responses (from March 2024), assuming the goal 3 eV spectral resolution of the X-IFU instrument. The result is shown in Fig. \ref{NewAthena_sim}. Even the brief 60 ks observation is sufficient to detect the outflow at a very high significance, with a fit improvement of \delcstat$~>1500$ over the baseline continuum. Thus, with \athena\, it will be possible to detect the outflow and characterize it in a single pointing, and track its evolution with time.

In the last part of this analysis, we explored the spectral residuals in the simultaneous EPIC datasets of GSN 069, previously found in the $0.6-0.8$ keV energy range by \citet{Miniutti+23a}. We confirm these residuals are present during both quiescent states, and find that they can be reproduced by an outflow of the same kinematic properties as the one detected in the RGS data, but likely with a much lower ionization parameter. This would indicate a multi-phase nature of the outflow, one phase producing spectral features more strongly in the RGS data, while the other dominating in the EPIC band. Alternatively, the residuals may not originate entirely from the outflow seen in the RGS, but instead could include contributions from a highly blueshifted ultra-fast outflow in absorption whose broadened spectral features cannot be detected with the RGS because of lack of signal-to-noise. This highly broadened UFO would have a projected velocity of about 0.2-0.3c and hence be kinematically unrelated to the outflow seen in the RGS. 

On the other hand, curiously, the QPE state EPIC spectra do not need any extra ionized absorption to reproduce the observed residuals. The outflow seen in RGS completely reproduces these features.


At this stage, both solutions for the EPIC residuals are plausible and further analysis of the combined RGS and EPIC dataset is needed. In this initial analysis, we ignored any long-term variability of GSN 069 since 2010 by stacking the data simply by spectral state. However, any variability in the Wien tail of the blackbody, or variability in the ionization and column density of the ionized outflow could bias the interpretation of the EPIC residual, which is very broad but located where EPIC has very limited spectral resolving power. In any case, both hypotheses indicate a likely complex ionization, and possibly also kinematic structure of outflowing plasma in the vicinity of GSN 069.

\subsection{Comparison between the states of GSN 069 and with TDE ionized outflows}


By splitting the RGS dataset, we are able to track the evolution of the plasma properties during the different spectral states of GSN 069. The plasma column density appears to be consistently around $1\times10^{22}$ \pcm\ in both quiescent states, comparable with the outflow detected in the TDE ASASSN-14li \citep{Miller+15}. The column density of the absorber during the QPE state is fully consistent with that during the Quiescence with QPEs. However, no emission component is detected in the QPE state, with an upper limit many times below the best-fitting column density or solid angle from the analysis of Quiescence with QPEs.

The ionization parameter of the outflow is quite high at around \logxi$\sim4$, again very similar to the outflow detected in ASASSN-14li. However, the outflow of GSN 069 is much faster, as the observed velocity of the blueshifted absorber in ASASSN-14li was only about 300 km/s \citep{Miller+15}. ASASSN-20qc, another TDE with an X-ray detected outflow \citep{Kosec+23b}, shows an outflow with a lower ionization of \logxi$\sim3$ (plus a less ionized second component). The velocity of the outflow in ASASSN-20qc is higher than in ASASSN-14li at 900 km/s, but even this is still much slower than what we observe in GSN 069. 

Importantly, neither of these TDE outflows show as strong line emission components as GSN 069. \citet{Miller+15} noted potential evidence for a weaker ionized emission component in ASASSN-14li, while no such emission was detected in ASASSN-20qc. 

The velocity of the ionized emitter in GSN 069 (up to 3000 km/s) is surprising as it is comparable with the projected velocity of the blueshifted absorber. In a scenario where the outflow is axisymmetric, one would expect the centroid of the emission line to be roughly stationary rather than significantly redshifted, except part of its blue wing would be affected by the line-of-sight absorption. This would apply regardless of the emitter solid angle $\Omega$ (which we measured here using line strengths rather than using plasma velocity structure). The redshift of the emitter could therefore indicate that the velocity structure of the outflow is non-axisymmetric. At this time it is unclear how this could occur in a QPE system. Alternatively, the velocity structure of the absorbing material could be much more complex, and affect the blue wing of the emitting plasma more heavily than currently modeled in the X-ray spectra. Unfortunately, this is difficult to probe with the current data quality. A second possible alternative is that the outflowing plasma is affected by gravitational redshift. However, this constraint would place the outflow only $\sim100$ R$_{\rm G}$ from the black hole, which is at odds with other outflow location estimates discussed below. Finally, the true outflow velocity could be significantly higher than apparent from its blueshift due to projection effects. However, in this case we would expect the ionized emitter to have a significantly higher velocity width, roughly comparable with the true outflow velocity, and some velocity structure non-axisymmetry would still be required within the outflow. Ultimately, further observations of GSN 069 are required to improve the data quality and understand this velocity structure discrepancy.

Intriguingly, the ionization parameter \logxi\ of the outflow is higher in the Quiescence with QPEs and in QPE state than during Quiescence without QPEs. Using the coupled spectral analysis results, we observe an increase in ionization parameter of $\Delta$\logxi$\sim0.4$. Assuming this increase is purely due to a change in ionizing luminosity, this would correspond to an increase in luminosity by a factor of 2.5. \citet{Miniutti+23a} showed that the power emitted in the QPE is only a very small fraction (up to 15\%) of the total bolometric luminosity of GSN 069 (most of the power emitted by the disk blackbody component is in the extreme UV), making it unlikely to cause this \logxi\ increase alone. However, we also note that during the QPEs, the SED of GSN 069 significantly hardens compared with both quiescent states. The QPE variability amplitude is much higher in the X-ray band compared with the bolometric value. Taking the best-fitting luminosity above 0.3 keV of each component from our spectral analysis, the flare-averaged QPE amplitude is about 3, but rises to 13 when comparing just the luminosities above 0.5 keV, the photon energy needed to excite the strongest observed outflow transition (N VII Ly$\alpha$). These SED shape variations can thus directly explain the observed higher ionization state of the outflowing plasma, despite little variability in the bolometric luminosity of GSN 069.

Curiously, we do not detect any ionized line emission during the QPE state, and place deep upper limits on its presence. The redshifted ionized emitter as seen during the Quiescence with QPEs is rejected by the QPE state data, assuming that it sees the same SED as we observe during the QPEs. This non-detection may have a number of explanations. The ionized emission could weaken or disappear if the emitting plasma is over-ionized. However, it is not clear why only the emitting plasma would over-ionize without the same happening to the absorbing plasma (which has a reasonably well constrained ionization parameter, Tables \ref{PION_results_decoupled} and \ref{PION_results_coupled}). 

Instead it is possible that the outflow is located so far away from GSN 069, that the light travel time from the black hole to the outflow is much longer than the time delay between consecutive QPEs. In this case, any variability in the illumination of the emission component (by the QPEs) would be washed out due to the time travel delays, and the SED that this emitter observes would be the weighted time average of the QPE and Quiescence with QPEs SEDs. This is not the case for the ionized absorption, which sees the immediate increase in the illuminating X-ray flux during the QPE as the absorber is located directly along our line of sight towards the ionizing source. In such a case, the contrast of the emission lines against the source continuum would be much lower than during the Quiescence with QPEs because the QPE continuum is much stronger than the weighted average of the QPE and Quiescence with QPEs continua. At the same time, this would likely weaken the upper limit we obtained on the presence of ionized emission (where we assumed that the emitter observes the QPE SED).

An alternative explanation is that the ionized emission is located relatively close to the black hole, but responds with a delay due to light travel time of more than $1-2$ ks (the duration of the QPE). In that case, the response of the emitter to the QPE would be extracted in the spectrum of Quiescence with QPEs, rather than directly in the QPE spectrum (which only contains the QPE itself). This hypothesis could be tested by splitting the Quiescence with QPEs spectrum into two halves by time elapsed since the previous QPE. Unfortunately, we do not have the data quality to achieve this with the current \xmm\ dataset. Further observations, specifically of the QPE active state, are required.

We do not make any conclusions on the measurement of plasma velocity widths. Since our spectra are composed from large stacks of many individual observations, any evolution in the outflow velocity (over the timescale of several years) would imprint on these stacks by broadening the spectral lines. For the same reason, it is difficult to draw strong conclusions about the best-fitting outflow velocity and its evolution with GSN 069 state. We consider the outflow velocity generally consistent across the different states.

Finally, we also tested for non-Solar abundances in the outflow by fitting for the abundance of N, which in our dataset has the highest signal-to-noise elemental line (N VII). N appears strongly over-abundant with an abundance (with respect to other metals) of $10-50$. This is a rather high value, but we do not consider this result abnormal as it is in line with the UV spectroscopy of GSN 069 by \citet{Sheng+21}, who found an N overabundance of at least 10. Similar unusual abundances were also found in 3 optical and UV-detected TDE outflows \citep{Cenko+16, Kochanek+16, Yang+17}, again indicating an overabundance of N by at least a factor of 10. The most extreme of these cases is the TDE ASASSN-14li, with an N/C overabundance of at least 300 in the X-ray detected outflow \citep{Miller+23}.

For TDEs, such unusual abundances may indicate disruption of massive, or stripped stars \citep{Miller+23, Mockler+24}. In the former case, N/C overabundances of $\sim10$ may be reached as the core of the star is enriched with excess N, revealed and accreted as the star is entirely disrupted. Even higher N over-abundances (N/C$>100$) can be achieved through the disruption of a stripped star, a result of stellar binary interaction, because its outer layers are disproportionately enriched with N \citep{Mockler+24}. For GSN 069, the explanation may be analogous. While there is currently no definitive proof that GSN 069 underwent a TDE, its long-term lightcurve strongly resembles that of a partial TDE \citep{Miniutti+23a}. Additionally, one of the other QPE systems, AT2019qiz, began its transient activity as an optically-selected TDE \citep{Nicholl+24}, and another TDE, AT2019vcb, is a QPE system candidate \citep{Quintin+23}. Hence, there is connection between TDEs and at least some QPE systems. The UV and X-ray measurements of N over-abundance thus contribute to the growing evidence that the recent activity of GSN 069 began with a (likely partial) TDE, and the disrupted star was a massive or a stripped star.

\subsection{Outflow location and energetics}

As this is the first detection of an outflow in a QPE source, we establish constraints on its location and energetics. We note that below we only calculate these quantities for the outflow confirmed via the RGS spectra, not the potential UFO seen in the EPIC data. We assume that the mass of the black hole in GSN 069 is $10^{6}$ M$_\odot$ \citep{Miniutti+23a}.

First, we put an upper limit on the distance of the outflow from the ionizing source. This is given by the ionizing balance of plasma, from the definition of the ionization parameter $\xi=L_{\rm{ion}}/(nR^{2})$ (where $L_{\rm{ion}}$ is the 13.6 eV--13.6 keV ionizing luminosity, $n$ the outflow number density, and $R$ is the distance from the ionizing source) and the definition of the column density $N_{\rm{H}}=n\Delta R$ (where $\Delta R$ is the thickness of the outflow layer):

\begin{equation}
    R = \frac{L_{\rm{ion}}}{N_{\rm{H}}\xi} \frac{\Delta R}{R}
\end{equation}

By assuming the maximum thickness of the absorbing layer ($\Delta R/R=1$), we obtain the maximum distance of the outflow from the ionizing source. We take the best-fitting values from the coupled absorption-emission spectral analysis. The maximum distance derived from Quiescence without QPEs is $9\times10^{16}$ cm ($6\times10^5$ $R_{\rm{G}}$ for a $10^6$ M$_\odot$ black hole, 0.03 pc), while the Quiescence with QPEs provides a tighter upper limit of $2\times10^{16}$ cm ($1\times10^5$ $R_{\rm{G}}$, 0.006 pc). Clearly, even the tighter of these two limits is not very constraining \citep[for illustration, the inferred blackbody emitting radius of the QPEs themselves is about $10^5$ times smaller,][]{Miniutti+23a}, but the result shows that we cannot be observing a galaxy-scale outflow.

By taking the projected absorber outflow velocity, $\sim$2600 km/s, and assuming that it is comparable to the escape velocity at these radii, we can obtain a very rough estimate for the outflow location. This value can be derived as follows:

\begin{equation}
    R = \frac{2GM}{v^2} = 2\frac{c^2}{v^2} R_{\rm G}
\end{equation}

\noindent which is about $3\times10^4$ $R_{\rm{G}}$, several times lower than its maximum distance from the ionizing source from the ionizing balance. However, we note that due to the projection of the outflow velocity into our line of sight, the true distance of the outflow from the black hole may be lower than this value. At the same time, its distance could also be higher if the outflow velocity is hyperbolic.

For the ionized emission to respond to QPEs (without washing out the QPE signal), the light travel time from GSN 069 to the ionized outflow must be shorter than the typical recurrence time between the QPEs. The time between consecutive QPEs (when in the QPE active state) is about 30 ks \citep{Miniutti+23a}, corresponding to a distance of about $6\times10^3$ $R_{\rm{G}}$. From the observed outflow velocity alone, it seems unlikely that the outflow (in emission) could respond to the QPEs coherently.

Additionally, further constraints on the plasma location can be placed by considering the response of the outflow in absorption to the QPEs. The absorbing plasma sees the same SED that we observe, with no time delays as is the case for the emitting plasma. If it has sufficient number density, it should react to the increase in the ionizing flux during the QPE and adjust its ionization parameter \citep{Krolik+95}. This is not observed in our study, as the ionization parameter does not appear to vary between the QPE and the Quiescence with QPEs. We stress that given the current data quality of the QPE state spectrum, this is not a strong constraint, but it will become a crucial probe of the outflow properties and location with higher quality X-ray observations, especially those with future X-ray observatories such as \athena\ \citep{Nandra+13}, \textit{Line Emission Mapper} \citep{Kraft+22} or \textit{Arcus} \citep{Smith+20}. With higher-quality data, these sources will be perfect for the application of time-dependent photoionization models such as \textsc{tpho} \citep{Rogantini+22} considering the large and fast variation of X-ray flux during the QPEs, and its effect on ionized plasma.

If the absorber does not respond during the QPE, its density must be so low that the ionization and recombination time of observed spectral line transitions is longer than the QPE itself. Taking the strongest observed N VII transition, we used the \textsc{ascdump} feature of the \textsc{pion} model to determine the recombination time of N VII as a function of number density. For simplicity, taking the recombination time to be at least 2000 s (roughly the duration of a QPE), we obtain an upper limit on the outflow density of $10^{7}$ cm$^{-3}$. Applying this limit to the ionizing balance equation, we obtain a lower limit for the distance of the outflow from the black hole of $5\times10^{15}$ cm (0.002 pc, $3\times10^4$ $R_{\rm{G}}$), fully consistent with the estimate we made using the outflow velocity argument, and with the upper limit from ionization balance.

Another relevant consideration is the flight time of the ionized outflow, and how it could relate to GSN 069 and its recent transient history. Assuming the outflow originates very close to the black hole, at the average projected velocity of 2600 km/s, it will reach the distance of $6\times10^3$ $R_{\rm{G}}$ in about $3\times10^6$ s (40 days), the distance of $3\times10^4$ $R_{\rm{G}}$ in 200 days and the distance of $1\times10^5$ $R_{\rm{G}}$ in 700 days. Even if the outflow does not decelerate over time (which is likely), its flight time to the current location is shorter than the time elapsed since the likely TDE event occurred in this source back in 2010. Hence it is unlikely that the outflow could originate from previous possible AGN activity of GSN 069 (before 2010) and be a remnant of that activity. The outflow is thus solidly associated with the recent nuclear transient behavior of GSN 069.

Finally, we calculate the mass outflow rate and the kinetic power of the outflow, assuming that it is continuously being launched by GSN 069. In these calculations, we follow the steps of \citet{Kosec+20a} in their Section 5.3. The mass outflow rate can be estimated via two approaches. The first one uses the ionizing balance of the outflow, and $M_{\rm{out}}$ can be obtained as follows:

\begin{equation}
\label{eqfinalMout}
\dot{M}_{\rm{out}} = 4 \pi \mu m_{\rm{p}} v \frac{L_{\rm{ion}}}{\xi}  C_{\rm{V}}  \frac{\Omega}{4 \pi}
\end{equation}

where $\mu$ is the mean atomic mass (assumed 1.2 for Solar abundances), $m_{\rm{p}}$ is the proton mass and $C_{\rm{V}}$ is the volume filling factor (unknown). This results in a mass outflow rate of $0.03-0.08$ $M_{\odot}~$/year (for Quiescence with and without QPEs, respectively). We note that there are two caveats to this calculation. First, we did not factor in the (unknown) value of $C_{\rm{V}}$, which will decrease the estimate. Secondly, the true value of the outflow velocity is higher than its line-of-sight projection (which we used here), which will increase the estimate. These caveats apply to the following calculations as well. The second approach makes the assumption that the observed projected outflow velocity is equal to the escape velocity.

\begin{equation}
\dot{M}_{\rm{out}} = 8 \pi \mu m_{\rm{p}} GM \frac{N_{\textrm{H}}}{v} \frac{1}{\delta R}  C_{\rm{V}}  \frac{\Omega}{4 \pi}
\end{equation}

where $G$ is the gravitational constant, $M$ is the mass of the black hole and $\delta R=\Delta R/R$ is the relative thickness of the outflow layer. Taking the most extreme case of $\Delta R/R=1$ again, we find a lower limit on the mass outflow rate (neglecting the $C_{\rm{V}}$ factor again) of $(3-7)\times10^{-3}$ $M_{\odot}~$/year. If the outflow persisted continuously until 2023 since the transient in GSN 069 began in July of 2010 (for 13 years) and its volume filling factor is high, the total ejected mass should be at least 0.04 $M_{\odot}$, but more realistically around $0.3-1.0$ $M_{\odot}$.

Therefore, if the transient in GSN 069 began by a tidal disruption of a $\sim M_{\odot}$ star, a significant fraction of its mass could have been ejected via this wind over the past 13 years. However, we stress we did not account for the volume filling factor of the plasma, which is unknown and could be much lower than 1, thus reducing the total mass in the outflow.

Finally, the kinetic power of the outflow is simply:

\begin{equation}
\dot{E}_{\rm{kin}}=\frac{1}{2} \dot{M}_{\rm{out}} v^2
\end{equation}

We then obtain $\dot{E}_{\rm{kin}} \sim 7\times 10^{39} - 2\times 10^{41}$ erg/s, about $6\times10^{-5}-1\times10^{-3}$ of the Eddington luminosity for a $10^6$ $M_{\odot}$ black hole. Therefore, even if this outflow persisted for a long amount of time, it is unlikely that it could significantly influence the host galaxy of GSN 069 \citep{diMatteo+05}.

\section{Conclusions} \label{sec:conclusions}

We performed the first in-depth analysis of high-resolution X-ray spectra of a quasiperiodically erupting source, leveraging the nearly 2 Ms archive of \xmm\ observations on GSN 069. We analyze the RGS grating spectra, first by stacking all the individual observations in a single dataset, and secondly in 3 dataset stacks based on the current states of GSN 069: Quiescence without QPEs, Quiescence with QPEs or QPE state. Our findings can be summarized as follows:

\begin{itemize}

\item The stacked RGS spectrum reveals an array of absorption and emission lines. The strongest feature, with a statistical significance far exceeding $4\sigma$, is located near the rest-frame transition of N VII. If associated with N VII, we observe a blueshifted absorption component at a velocity of $3500_{-600}^{+400}$ km/s, and a redshifted emission component with a velocity of $2800_{-1000}^{+500} $ km/s, very reminiscent of the P-Cygni line shape.

\item Ionized absorption is present in all states of GSN 069, with a projected outflow velocity of $1700-2900$ km/s. The best-fitting column density is $(0.9-1.7) \times 10^{22}$ \pcm, consistent with being stable across all 3 states. The outflow is highly ionized, with an ionization parameter \logxi\ of $3.9-4.1$ during Quiescence without QPEs, and \logxi\ of $4.5-4.6$ during Quiescence with QPEs and QPE states.

\item Ionized emission has a redshift of up to 2900 km/s, depending on the GSN 069 state. It is strongly detected in both quiescent states, with a column density and ionization parameter similar to the best-fitting values of the ionized absorption. This emission component can be interpreted as the same plasma seen in absorption, only located out of our line of sight. In that case, the outflow is consistent with covering the full $4\pi$ sky from the point of view of GSN 069.

\item No ionized line emission is observed during the QPEs, and we place deep upper limits on its presence, several $\sigma$ below the best-fitting values during the Quiescence with QPEs. This non-detection may be explained if the outflow is located sufficiently far from the ionizing source that the light-travel time significantly delays, or washes out any ionization response to the QPEs in the ionized emission.

\item We find strong evidence that nitrogen is highly over-abundant in the outflow compared with other metals. The best-fitting N abundance is $24^{+17}_{-10}$, consistent with a previous abundance study of GSN 069 using UV spectra \citep{Sheng+21}.

\item We place limits on the outflow location and energetics. It must be located at most 0.03 pc from the black hole ($10^5$ $R_{\rm{G}}$ assuming a $10^6$ M$_\odot$ black hole), with a travel time from the ionizing source of $\leq2$ years, and so cannot be a remnant outflow launched by previous black hole activity. It is hence directly related to the recent transient phenomena occurring in GSN 069 since 2010. 

\item Assuming the outflow is continuously launched, its mass outflow rate is $3\times10^{-3}-8\times10^{-2}$ $M_{\odot}~$/year, with a total ejected mass of up to 1~$M_{\odot}$ since 2010 (when the GSN 069 transient activity apparently began). Its kinetic power is $7\times 10^{39} - 2\times 10^{41}$ erg/s, and so the outflow does not have enough power to provide sufficient feedback to affect the evolution of the host galaxy of GSN 069.

\end{itemize}


\begin{acknowledgments}
 Support for this work was provided by the National Aeronautics and Space Administration through the Smithsonian Astrophysical Observatory (SAO) contract SV3-73016 to MIT for Support of the Chandra X-Ray Center and Science Instruments. Support for this work was provided by NASA through the NASA Hubble Fellowship grant HST-HF2-51534.001-A awarded by the Space Telescope Science Institute, which is operated by the Association of Universities for Research in Astronomy, Incorporated, under NASA contract NAS5-26555. CP acknowledges support by INAF Large Grant BLOSSOM and PRIN MUR 2022 SEAWIND (funded by NextGenerationEU). GM thanks grant n. PID2020-115325GB-C31 funded by MCIN/AEI/10.13039/50110001103 for support. MG is supported by the ``Programa de Atracci\'on de Talento'' of the Comunidad de Madrid, grant numbers 2018-T1/TIC-11733 and 2022-5A/TIC-24235.
\end{acknowledgments}

%

\vspace{5mm}
\facilities{\xmm
}


\software{SPEX \citep{Kaastra+96}, Veusz
          }




 \appendix

\section{Further checks of \xmm\ RGS observational background}
\label{app:RGSbkg}

Since GSN 069 is a very faint X-ray source, particularly in quiescence, its source flux is below RGS background level throughout most or all of the wavelength range, depending on the current source state. In Section \ref{sec:data} we performed a standard RGS data reduction using observational backgrounds, which works reasonably well throughout most of the wavelength range. The only exception are RGS data between 31 and 33.5 \AA\ that are strongly background-dominated, and so are excluded from this analysis. Our primary conclusion about the RGS background subtraction being reasonably accurate despite the high background levels is based on two sanity checks. First, in no RGS spectra analyzed in this work we observed large clusters of background-subtracted source data points with negative flux values, which would indicate inaccurate background subtraction. Second, in all spectral fits we employed a cross-calibration constant to account for any residual calibration differences between RGS 1 and 2 instruments. The value of this constant was always within 10\% of unity in our analysis, confirming that the source continuum level is very comparable in both instruments and so the background subtraction is unlikely to be systematically inaccurate. Nevertheless, below we perform two additional checks of RGS background subtraction.

\begin{figure*}
\begin{center}
\includegraphics[width=\textwidth]{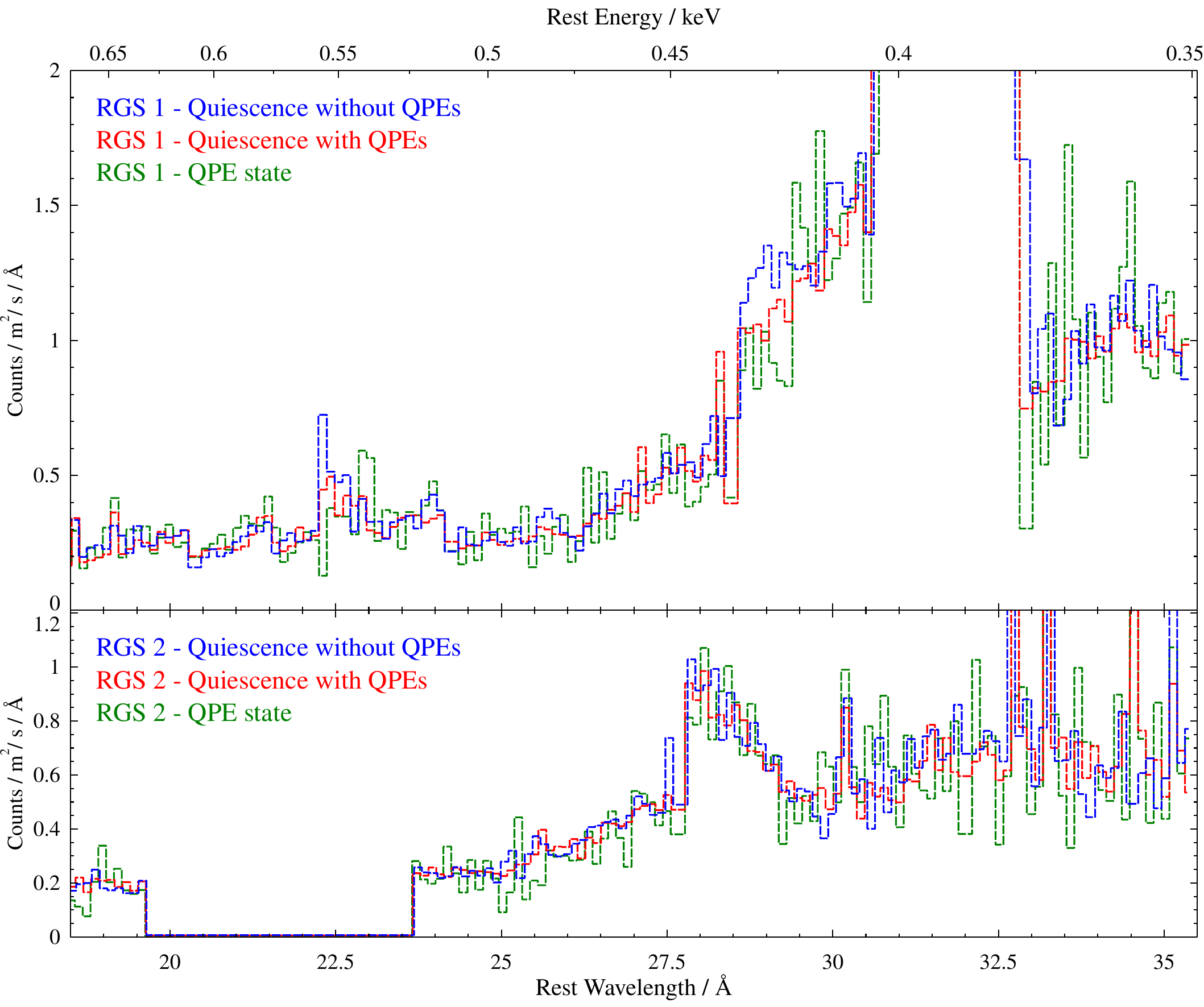}
\caption{Top panel: observational background of RGS 1 during Quiescence without QPEs (blue), Quiescence with QPEs (red) and the QPE state (green). Lower panel: observational background of RGS 2 (same color scheme used). We note that the Y-axis units are the same in both panels, showing that RGS 1 experiences overall higher background levels, particularly between 30.5 and 32.9 \AA\ (in the GSN 069 rest-frame). \label{RGS_bkg_comparison}}
\end{center}
\end{figure*}

First, we extract observational RGS backgrounds for the 3 different GSN 069 states. These backgrounds were extracted at 3 different (non-overlapping) times and so can serve as a comparison of the RGS background level versus wavelength over time, and show that the background level is stable. This is illustrated in Fig. \ref{RGS_bkg_comparison}. The figure shows that the background remains comparable during all three periods of time. While some small differences occur, they do not appear strong and systematic. The QPE state background shows the most noise with wavelength, which is expected because it has the shortest total exposure time (many times lower than the other two datasets). Additionally, the figure illustrates why we ignored the RGS 1 data between 30.5 and 32.9 \AA\ due to a step jump in the background level at these wavelengths.

\begin{figure*}
\begin{center}
\includegraphics[width=\textwidth]{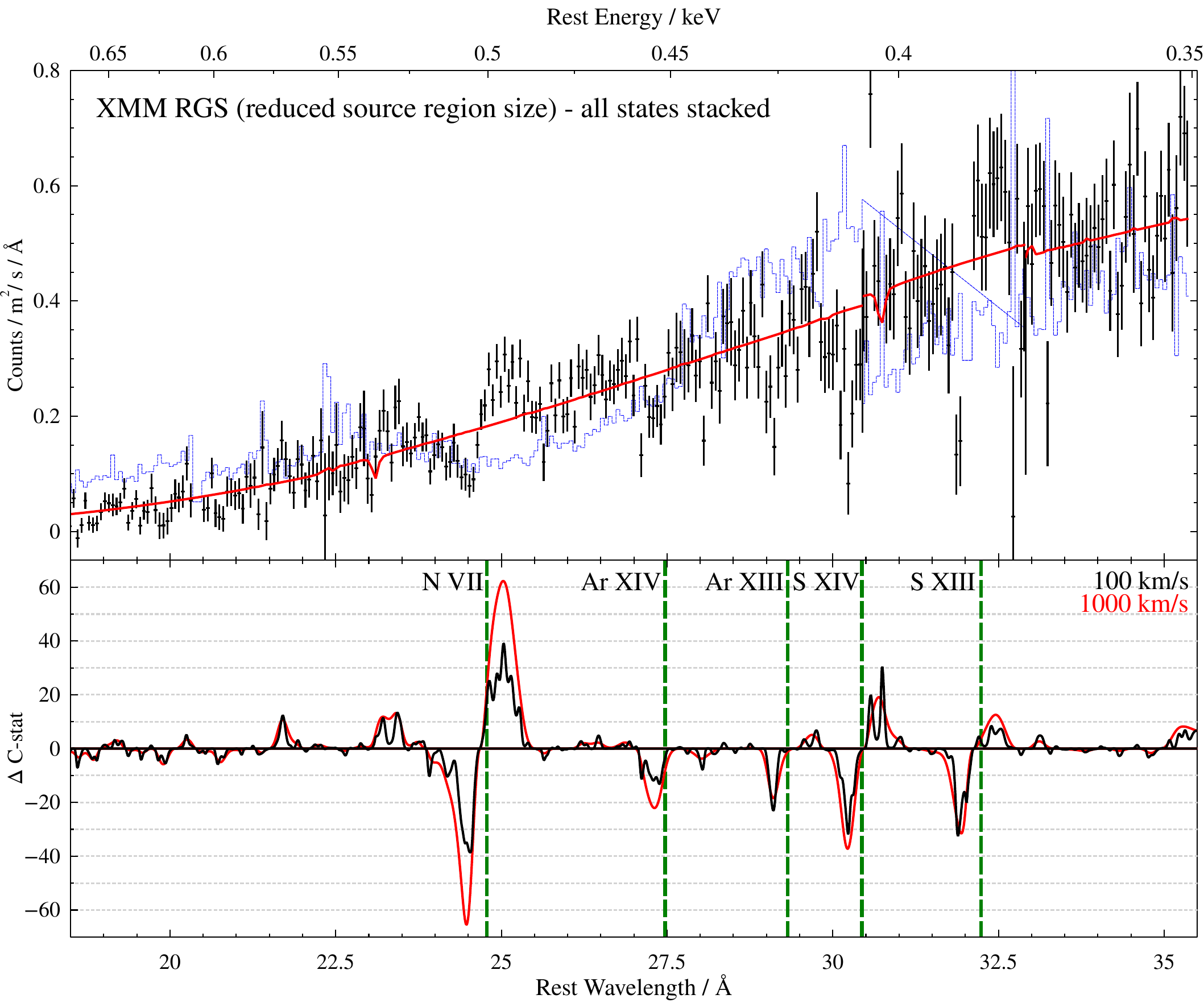}
\caption{Top panel: RGS grating spectrum of GSN 069, obtained by combining all available \xmm\ observations. The RGS source extraction region size was decreased to improve the signal-to-noise ratio. Most of the wavelength range contains RGS 1+2 data, stacked and overbinned for visual purposes only, except the range between 30.5 and 32.9 \AA\ (in GSN 069 rest-frame) which only contains RGS 2 data. The best-fitting continuum model is shown in red and the observational background is in blue. Bottom panel: Gaussian line scan of the same spectrum, with a line width of either 100 km/s (black) or 1000 km/s (red). Green dashed lines indicate the rest-frame wavelengths of transitions which can plausibly explain the strongest residuals.  \label{RGS_bkg_adjusted}}
\end{center}
\end{figure*}

Secondly, we perform a non-standard RGS data reduction to reduce the level of background compared to source flux, at the cost of a fraction of source counts. To achieve this, we adjust the \textsc{xpsfincl} keyword in the \textsc{rgsproc} reduction script that controls the source extraction region size. By significantly reducing the source region size, we increase the ratio of source to background flux because the background rate per area is roughly constant while the source counts are centrally peaked (GSN 069 is a point source). The value of \textsc{xpsfincl} roughly corresponds to the fraction of the RGS point spread function (PSF) enclosed by the source region. We tried a range of different values but here we show specifically the results with \textsc{xpsfincl}=80. For illustration, in a standard RGS reduction, typically a value of 90 or 95 is used. The analysis is performed on the total RGS stacked spectrum analyzed in Section \ref{sec:Gaussianscan}. The adjusted reduction procedure decreased the overall subtracted background count rate by a factor of 2.5 and the reduced total source count rate by about 10\%. The resulting stacked RGS spectrum is shown in Fig. \ref{RGS_bkg_adjusted}.

By comparing Figs. \ref{RGS_Gaussian_scan} and \ref{RGS_bkg_adjusted}, we can see that despite the large shift in the background level, the same strong residuals in the source spectrum remain. To visualize these features more clearly, we repeat the same Gaussian scan performed on the RGS data obtained with standard reduction scripts, and show it in the lower panel of Fig. \ref{RGS_bkg_adjusted}. The absolute \delcstat\ values of the individual residuals slightly shift, as the total statistics of this spectrum is somewhat different. However, qualitatively, we observe absorption and emission residuals at the exact same positions as in Fig. \ref{RGS_Gaussian_scan}, despite the overall background level (and the associated amount of background subtraction) changing by a factor of 2.5. Most spectral residuals are not coincident with any significant spikes in the background level. One exception is the emission residual near the S XIV rest-frame transition, which is located in the vicinity of a spike in observational background, and so caution is required in its interpretation (Fig. \ref{RGS_bkg_adjusted}). We conclude that majority, if not all observed lines are real and not an artifact of inaccurate background subtraction.

Finally, we note that we still prefer to use the standard RGS data reduction for spectral fitting and particularly for state-resolved analysis, as it produces more total source counts, and additionally has been more thoroughly tested by previous works using RGS datasets over the past 25 years of the \xmm\ mission. Adjusting the source extraction region size keyword too far off the standard values could skew the source spectrum due to potential energy dependence of the instrument PSF. This is unlikely to affect the narrow lines themselves, but it could systematically bias the measured source continuum.

\section{Detailed information about \xmm\ observations}
\label{app:xmmdata}

In Table \ref{xmmdata} we list the details of \xmm\ observations used in this work. All of these observations were combined to create the full RGS stack, as well as the state-resolved stacked spectra.

\begin{deluxetable}{cccc}
\tablecaption{Details of the individual \xmm\ observations. We note that the duration is given before any flare filtering. \label{xmmdata}}
\tablewidth{0pt}
\tablehead{
\colhead{Observation ID} & \colhead{Start Date} & \colhead{Duration} & \colhead{QPE detection}  \\
\nocolhead{} & \nocolhead{} & \colhead{ks} & \colhead{}
}
\startdata 
0657820101 &  2010-12-02 &  14913 &  no \\
0740960101 &  2014-12-05 &  95100 &  no \\
0823680101 &  2018-12-24 &  63300 &  yes \\
0831790701 &  2019-01-16 & 141400 &  yes \\
0851180401 &  2019-05-31 & 135400 &  yes \\
0864330101 &  2020-01-10 & 141000 &  yes \\
0864330201 &  2020-05-28 & 133100 &  no \\
0864330301 &  2020-06-03 & 133200 &  yes \\
0864330401 &  2020-06-13 & 136100 &  no \\
0884970101 &  2021-06-30 &  53000 &  no\\
0884970201 &  2021-12-03 &  55300 &  no \\
0913990201 &  2022-07-07 &  59180 &  yes \\
0914790101 &  2022-11-30 &  19800 &  no* \\
0914790401 &  2022-12-02 &  15000 &  yes\\
0914790201 &  2022-12-04 &  21900 &  yes \\
0914790301 &  2022-12-08 &  17990 &  yes \\
0914790501 &  2022-12-11 &  15000 &  no* \\
0914790601 &  2022-12-13 &  15000 &  no* \\
0914791101 &  2022-12-14 &  18000 &  yes \\
0914790701 &  2022-12-16 &  14800 &  yes \\
0914790801 &  2022-12-18 &  13000 &  no* \\
0914790901 &  2022-12-20 &  16460 &  yes \\
0914791001 &  2022-12-22 &  20900 &  yes \\
0914791201 &  2022-12-26 &  15000 &  yes \\
0914791301 &  2022-12-30 &  18000 &  no* \\
0914791401 &  2023-01-05 &  16000 &  no* \\
0914791501 &  2023-01-08 &  13900 &  yes \\
0914791601 &  2023-01-10 &  13500 &  no* \\
0914791701 &  2023-01-12 &  29700 &  yes \\
0914791801 &  2023-01-14 &  13900 &  no* \\
0914791901 &  2023-01-15 &  13000 &  yes \\
0914792701 &  2023-05-29 &  69400 &  yes \\
0914792901 &  2023-06-09 & 119700 &  yes \\
0914793101 &  2023-07-09 & 108200 &  yes \\
\enddata
\tablecomments{
$^{*}$No QPEs were detected in these short exposure observations (briefer than the typical time interval between two QPEs), but we assumed that GSN 069 was constantly in the QPE active state between Nov 2022 and Jan 2023.}
\end{deluxetable}

\section{Systematic search for ionized absorption in GSN 069}
\label{app:systematic_search}

To verify that we identified the observed spectral residuals with correct elemental transitions, and accurately measured the plasma physical properties, we perform a systematic automated search for ionized absorbers in the 3 state-resolved RGS spectra. In this search we test that there are no ionized absorber solutions which work with the RGS data better than our interpretation, i.e. we establish whether our solutions are indeed the global best fits for these spectra using the \textsc{pion} photoionization model. To test this hypothesis, we fit (in an automated fashion) the different GSN 069 state RGS spectra with ionized absorbers of various physical properties, spanning a broad range of outflow velocities, ionization parameters and velocity widths. We note that we only search for ionized plasma in absorption, as performing such a search for plasma in emission is prohibitively computationally expensive using the \textsc{pion} model in \textsc{spex}.

We closely follow the systematic spectral search method adopted in Section 3.2 of \citet{Kosec+20b}, except here we apply the \textsc{pion} ionized absorption model instead of \textsc{xabs} applied in that work. We search all 3 GSN 069 state spectra: Quiescence without QPEs, Quiescence with QPEs and QPE state, and test ionized absorber properties as follows. Outflow velocities from +15000 km/s (infalling) to -100000 km/s (ultra-fast outflow), ionization parameters \logxi\ from 2.0 to 6.0 (mildly ionized to very highly ionized) as well as 3 different velocity widths: 250 km/s, 1000 km/s, 2500 km/s. Column density is a fitted parameter at each step of this grid and so it is not a grid parameter.

\begin{figure}
\begin{center}
\includegraphics[width=0.65\textwidth]{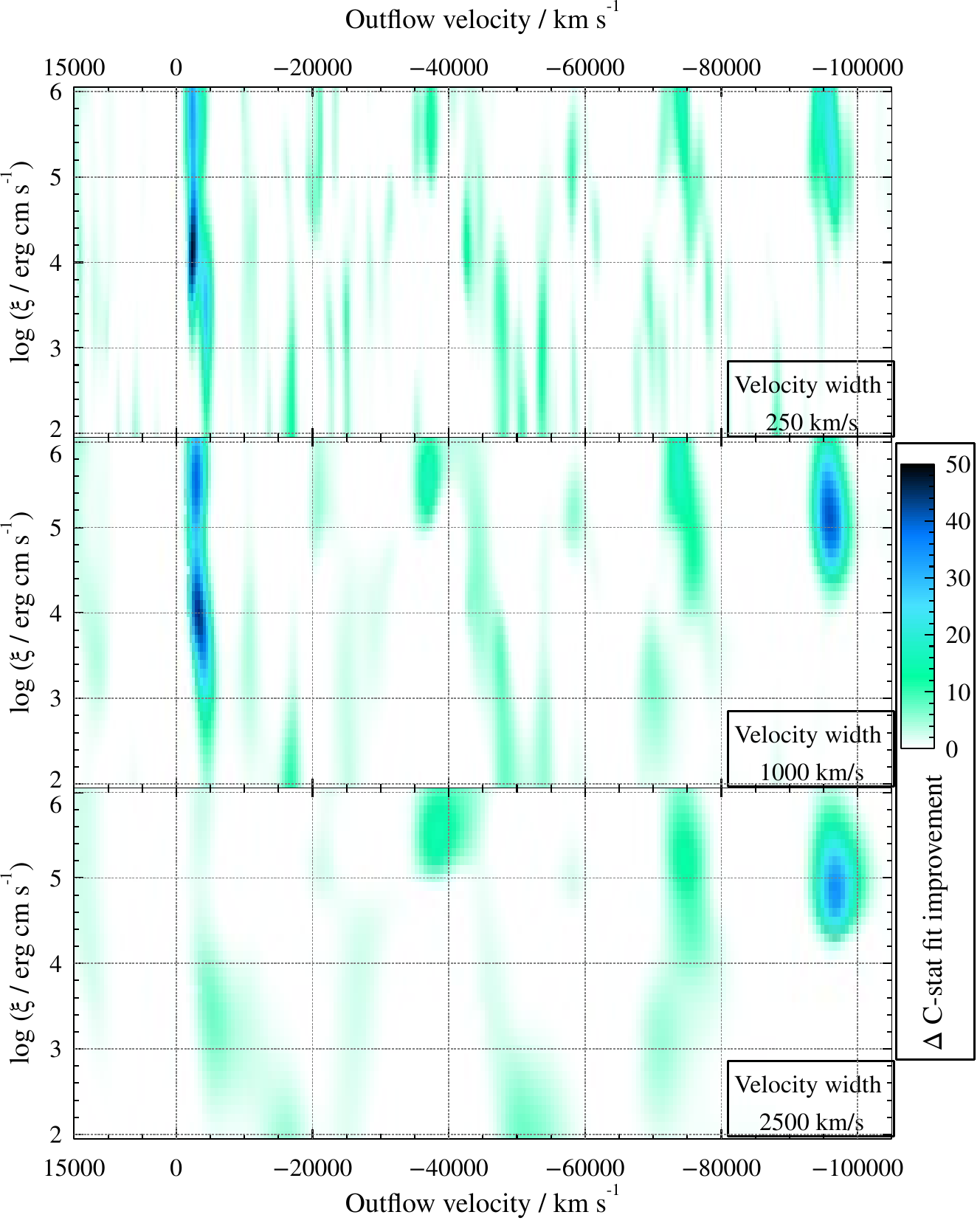}
\caption{Systematic search for ionized plasma in absorption of the Quiescent state without QPEs RGS spectrum. The top panel shows searches with a velocity width of 250 km/s, middle panel with a velocity width of 1000 km/s and the bottom panel with a velocity width of 2500 km/s. The X-axis of all panels is the outflow velocity, ranging from +15000 km/s (redshifted inflow) to -100000 km/s (high-velocity blueshifted outflow). The Y-axis is the ionization parameter of the \textsc{pion} component. The colour shows the fit improvement of adding an absorber of certain parameters to the baseline continuum, according to the colour scale on the right. \label{Search_noQPE}}
\end{center}
\end{figure}

\begin{figure}
\begin{center}
\includegraphics[width=0.65\columnwidth]{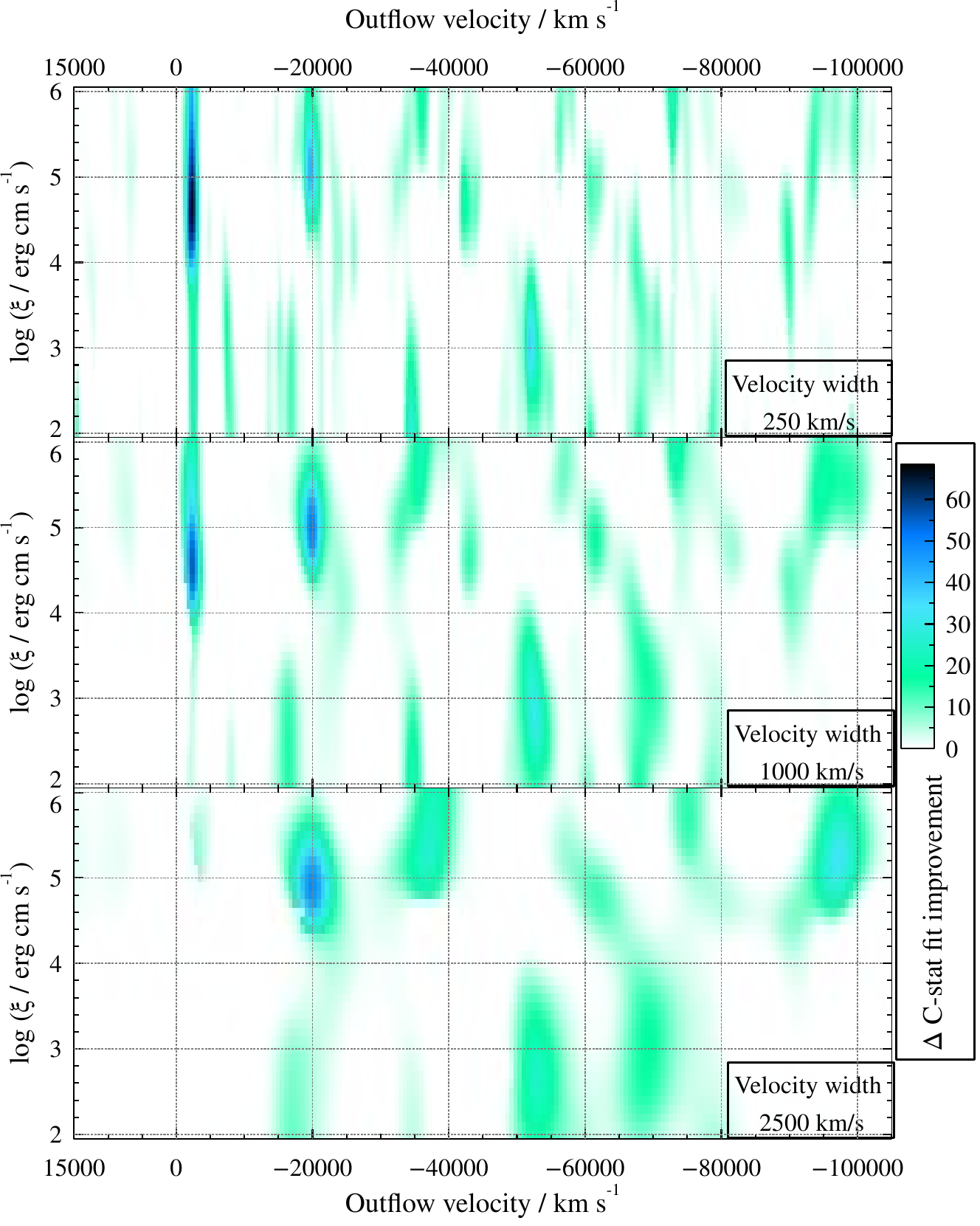}
\caption{Same as Fig. \ref{Search_noQPE}, but showing a systematic search of the Quiescent state with QPEs RGS spectrum. \label{Search_Quiescent}}
\end{center}
\end{figure}

\begin{figure}
\begin{center}
\includegraphics[width=0.65\columnwidth]{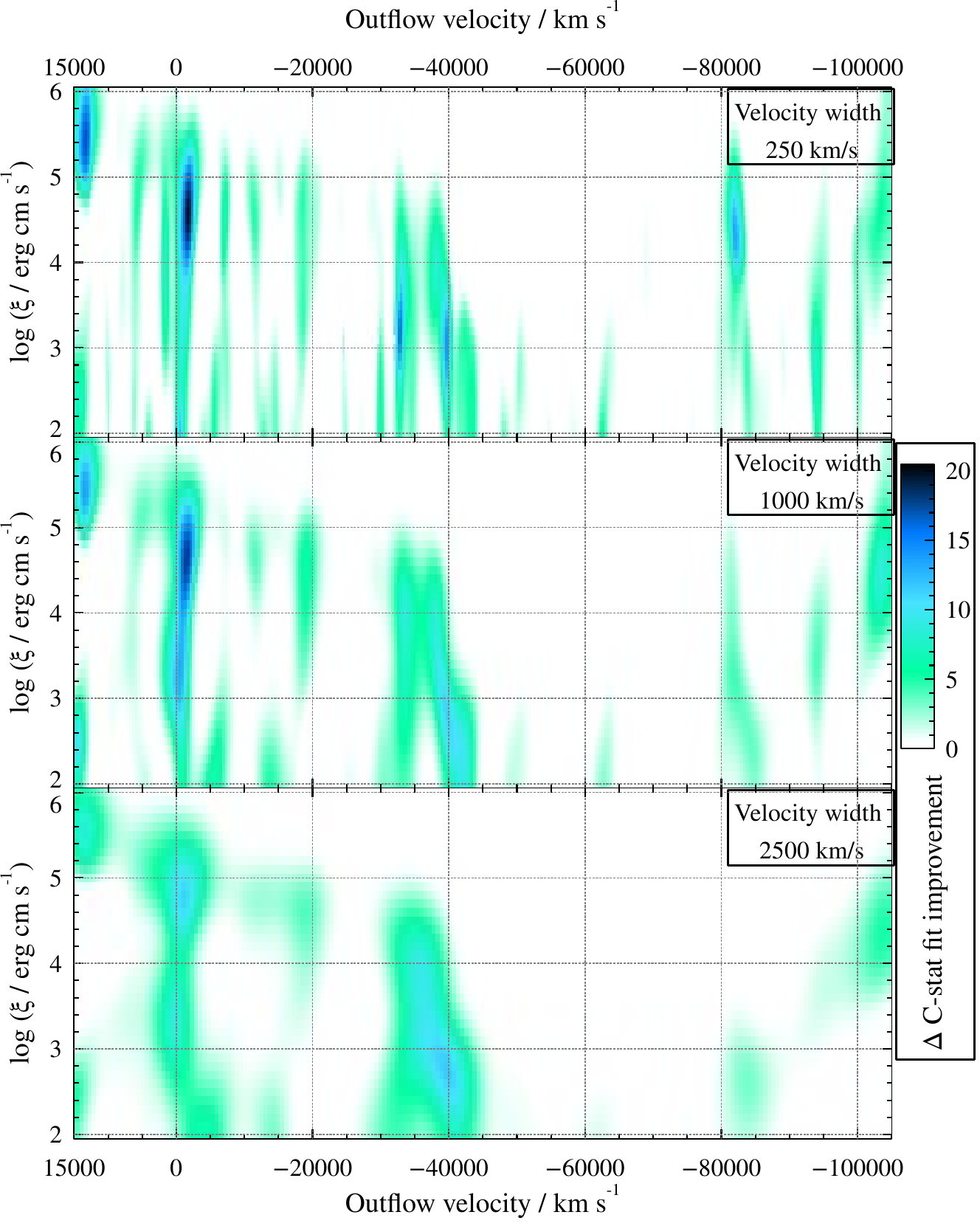}
\caption{Same as Fig. \ref{Search_noQPE}, but showing a systematic search of the QPE state RGS spectrum. \label{Search_Flare}}
\end{center}
\end{figure}

For each grid step, we fit the model containing the baseline continuum (\textsc{dbb} for Quiescence with or without QPEs and \textsc{dbb+bb} for the QPE state) plus the ionized absorber of the current grid step parameters and recover the \delcstat\ fit improvement in comparison with the baseline continuum fit. Performing the search over the full grid, we should obtain a \delcstat\ parameter space which shows a global maximum for certain ionized absorber parameters, which are the most preferred solution for this dataset. The results of this search are shown for Quiescence without QPEs in Fig. \ref{Search_noQPE}, for Quiescence with QPEs in Fig. \ref{Search_Quiescent} and for the QPE state in Fig. \ref{Search_Flare}.

We find that a highly ionized absorber, with \logxi\ of $4-5$, an outflow velocity of about 2500 km/s, and a velocity width of $250-1000$ km/s is the most preferred solution in all 3 states of GSN 069, resulting in the highest values of \delcstat. We also see secondary peaks in these systematic searches, indicating possible alternative solutions to the primary interpretation of the ionized absorber. For Quiescence without QPEs, this is at an outflow velocity of about -95000 km/s, with \logxi\ of 5 and a velocity width of 1000 km/s. For the Quiescence with QPEs, this is at an outflow velocity of -20000 km/s, with a \logxi\ of 5 and a velocity width of $1000-2500$ km/s. Finally, for the QPE state, there is a secondary solution with an inflow velocity of 13000 km/s, a \logxi\ of 5.5 and a velocity width of 250 km/s. However, we stress that all these secondary solutions are weaker (with lower \delcstat) in all 3 GSN 069 states than the preferred primary 2500 km/s solution. Importantly, none of the secondary solutions repeat among the different spectral states, while the primary solution remains consistently similar (and preferred) across all of the GSN 069 states. Therefore, we conclude that the best solution to the RGS dataset is a highly ionized outflow in absorption with a velocity of about 2500 km/s.


\bibliography{References}{}
\bibliographystyle{aasjournal}



\end{document}